\documentclass[aps,prc,twocolumn,amsmath,superscriptaddress,floatfix,nofootinbib]{revtex4}
\usepackage{graphicx,epsfig,epstopdf,longtable}
\newcommand{\beqy}{\begin{eqnarray}}
\newcommand{\eeqy}{\end{eqnarray}}
\newcommand{\bmlet}{\begin{subequations}}
\newcommand{\emlet}{\end{subequations}}

\begin{document}

\title{Unified equations of state for cold non-accreting neutron stars with 
Brussels-Montreal functionals.  II. Pasta phases in semi-classical 
approximation}

\author{J. M. Pearson}
\affiliation{D\'ept. de Physique, Universit\'e de Montr\'eal, Montr\'eal
(Qu\'ebec), H3C 3J7 Canada}

\author{N. Chamel}
\affiliation{Institut d'Astronomie et d'Astrophysique, CP-226, Universit\'e
Libre de Bruxelles, 1050 Brussels, Belgium}

\author{A. Y. Potekhin}
\affiliation{Ioffe Institute, Politekhnicheskaya 26,
194021 Saint Petersburg, Russia}

\date{\today}

\begin{abstract}

We generalize our earlier work on neutron stars, which assumed spherical 
Wigner-Seitz cells in the inner crust, to admit the possibility of pasta 
phases, i.e., non-spherical cell shapes. Full fourth-order extended 
Thomas-Fermi calculations using the density functional BSk24 are performed for 
cylindrical and plate-like cells. Unlike in our spherical-cell calculations we 
do not include shell and pairing corrections, but there are grounds for 
expecting these corrections for pasta to be significantly smaller. It is 
therefore meaningful to compare the ETF pasta results with the full 
spherical-cell results, i.e., with shell and pairing corrections included.
However, in view of the many previous studies in which shell and pairing 
corrections were omitted entirely, it is of interest to  compare our pasta 
results with the ETF part of the corresponding spherical calculations.
Making this latter comparison we find that as the density increases the
cell shapes pass through the usual sequence sphere $\rightarrow$ cylinder 
$\rightarrow$ plate before the transition to the homogeneous core. The filling
fractions found at the phase transitions are in close agreement with 
expectations based on the liquid-drop model. On the other hand,
when we compare with the full spherical-cell results, we find 
the sequence to be sphere $\rightarrow$ cylinder $\rightarrow$ sphere 
$\rightarrow$ cylinder $\rightarrow$ plate. In neither case do any
``inverted'', i.e., bubble-like, configurations appear. The analytic fitting
formulas for the equation of state and composition that we derived in our 
earlier work on the assumption of spherical cell shapes for the entire density 
range from the outer crust to the core of a neutron star are found to remain 
essentially unchanged for pasta shapes. Here, however, we provide more accurate
fitting formulas to all our essential numerical results for each of the three 
phases,
designed especially for the density range where the nonspherical shapes are 
expected, which enable one to capture not only the general behavior of the 
fitted functions, but also the differences between them in different phases.
\end{abstract}

\maketitle

\section{Introduction}
\label{intro}

Working within the framework of the theory of nuclear energy-density 
functionals, we have recently published calculations of the equation of state 
(EoS) and the composition of the ground state of neutron-star matter that are 
unified in the sense that all three major regions of these stars are treated 
using the same functional~\cite{pea18}.
The importance of such a unified treatment has been discussed, e.g., in 
Refs.~\cite{DouchinHaensel01,HP04,HPY,fant13,pot13,fort16}.

The outermost of the three regions, the ``outer crust'', consists, we recall, 
of an assembly of bound nuclei and electrons that globally is electrically 
neutral. The nuclei in this region become more and more neutron rich with 
increasing depth, until at a mean baryon number density $\bar{n}$ of around 
$2.6 \times 10^{-4}$ fm$^{-3}$ unbound neutrons appear. This so-called 
``neutron drip'' marks the transition to the ``inner crust'', an
inhomogeneous assembly of neutron-proton clusters and unbound neutrons, 
neutralized by electrons. As shown in Ref.~\cite{pea18}, at much higher 
densities a substantial amount of free protons may appear, which is called 
``proton drip'' because of the analogy (albeit incomplete) with the neutron 
drip. By the point where $\bar{n}$ has risen to about 0.08 fm$^{-3}$ the 
inhomogeneities have been smoothed out: this is the ``core'' of the star. 

The calculations of Ref.~\cite{pea18} were actually performed with four 
different functionals, each functional being used in all three regions of the
neutron star. These functionals, labeled BSk22, BSk24, BSk25, and BSk26, belong
to a family of functionals that have been developed not only for the study
of neutron-star structure but also for the general purpose of providing a 
unified treatment of a wide variety of phenomena associated with the birth and
death of neutron stars, such as supernova-core collapse and neutron-star 
mergers, along with the r-process of nucleosynthesis (Ref.~\cite{gcp13} and 
references therein). They are based on generalized Skyrme-type forces and 
density-dependent contact pairing forces, the formalism for which is presented 
in Refs.~\cite{cgp09,cha10}. The parameters of the functionals were determined 
primarily by fitting to essentially all the nuclear-mass data of the 2012 
Atomic Mass Evaluation~\cite{ame12}; we calculated nuclear masses using the 
Hartree-Fock-Bogoliubov (HFB) method, with axial deformation taken into 
account. In making these fits we imposed certain constraints, the most 
significant of which is to require consistency, up to the densities prevailing 
in neutron-star cores, with the EoS of homogeneous pure neutron matter, as 
calculated by many-body theory from realistic two- and three-nucleon forces. 

For the nuclear masses required for the treatment of the outer crust in 
Ref.~\cite{pea18} we took experimental values, where available, and otherwise 
used the values determined by
HFB calculations with the appropriate functional. In the interest of 
consistency it might appear desirable to apply this same method to the inner 
crust, but this would require unacceptably long computer times: see, e.g., 
Ref.~\cite{bc18} for a recent summary of the situation. Instead, we adopted the
ETFSI (fourth-order extended Thomas-Fermi plus Strutinsky integral) 
method~\cite{dut04,ons08,pcgd12,pcpg15}. It consists of a full ETF 
treatment of the kinetic-energy and spin-current densities, with shell 
corrections added perturbatively and pairing handled in the
Bardeen-Cooper-Schrieffer (BCS) approximation. 
It is to be noted that the errors incurred by the latter approximation lie 
within the errors of the ETFSI approach~\cite{pastore16}. 
The ETFSI+BCS method was originally developed for the treatment of bound 
nuclei~\cite{dut86}, and we discuss in Ref.~\cite{pea18} the high degree of
accuracy with which it approximates the HF+BCS method in that context. When
applied to the calculation of the EoS of the inner crust we take only the 
proton shell corrections into account. 
The neutron shell corrections have been shown to be much smaller than the 
proton shell corrections~\cite{oy94}, as might be expected, given that the 
spectrum of unbound neutron single-particle (s.p.) states is 
continuous~\cite{ch07,ch12}. We 
thus simply neglect the neutron shell corrections, an option that is not 
available in the HFB method but is possible in the ETFSI method because of its 
perturbative treatment of these corrections. We likewise include proton 
pairing~\cite{pcpg15} but not neutron pairing.  

Our ETFSI calculations of the inner crust have in all our previous work assumed
spherically-symmetric WS cells as an approximate description. Usually, our 
solutions are drop-like, i.e., the density is higher in the center of the cell 
than at the surface, but in Ref.~\cite{pea18} ``inverted'' solutions were found
at some points close to the interface with the core. In the literature, such 
``bubble-like'' solutions are often accompanied by nonspherical solutions at 
neighboring densities (see, for example, Ref.~\cite{pet95}). In the present 
paper we therefore go beyond the assumption of spherical WS cells.

Investigation of nonspherical configurations goes back to the work of 
Ravenhall {\it et al.}~\cite{rpw83} and Hashimoto {\it et al.}~\cite{hash84}, 
who considered, as an alternative to spherical shapes, infinitely long 
cylinders and plates of infinite extent. These are referred to as
``spaghetti'' and ``lasagna'' respectively, and thus collectively as ``pasta''.
{\bf Other early papers to be noted are those of Williams and Koonin~\cite{wk85}
and Lorenz et al~\cite{lrp93};} see also Refs.~\cite{pet95,bc18} for reviews. 
The part of the inner crust 
in which pasta shapes prevail is referred to as the ``mantle". It has been 
shown to behave like liquid crystals~\cite{PP98}, in contrast to the rest of 
the inner crust, which can be regarded as a solid. Early studies of the pasta 
phases were formulated in terms of the liquid-drop model, but even within the 
framework of energy-density functional theory the question as to whether or not
nonspherical shapes can ever be energetically favored is still a matter of some
controversy. For example, the second-order ETF calculations of Ref.~\cite{mu15}
with parametrized nucleon distributions show a transition to nonspherical 
shapes at a density $\bar{n}$ close to 0.06 fm$^{-3}$ in the inner crust for 
the functional SLy4, while the zeroth-order TF calculations of 
Ref.~\cite{vinas17} with the same functional yield no deviation from a 
spherical shape. 

We present in this paper what we believe to be the first full fourth-order ETF
calculations of pasta within the WS approach. However, of the four functionals 
considered in Ref.~\cite{pea18} we limit ourselves here to the phenomenologically 
superior BSk24 {\bf (see Ref.~\cite{gcp13})}. In Section~\ref{calc} we describe our 
method of calculating cylindrical and plate-like WS cells, while our results are 
summarized in Section~\ref{resu}. The conclusions will be found in 
Section~\ref{concl}. In Appendix~A we provide proofs of some equations used in 
the main text. Extended numerical results are presented as Supplemental 
Material in electronic form.

\section{Calculation of pasta phases}
\label{calc}

We draw here heavily on our earlier work on spherical WS 
cells~\cite{opp97,ons08,pcgd12,pea18}, emphasizing mainly the ways in which it
has to be modified for pasta calculations. For both the infinitely long 
cylindrically symmetric WS cells of the ``spaghetti'' phase and the  plate-like
WS cells of infinite extent representing the ``lasagna'' phase we still write 
for the total energy per nucleon 
\beqy\label{1}
e = e_\mathrm{Sky} + e_\mathrm{C} + e_{e} 
- Y_{p}\,Q_{n,\beta} \quad .
\eeqy
Here the first term denotes the total nuclear energy per nucleon corresponding
to our chosen Skyrme functional, the second and third terms the Coulomb and
electronic kinetic energy per nucleon, respectively, while the last term takes 
account of the neutron-proton mass difference, $Q_{n,\beta}$ (= 0.782 
MeV) being the beta-decay energy of the neutron. In this last term 
$Y_p$ is the proton fraction $Z/A$, $Z$ and 
$A$ being respectively the number of protons and nucleons in the WS cell for 
the spherical case, the number per unit length of cylindrical cells or the 
number per unit area of plate-like cells. (We have dropped for convenience the 
constant neutron mass term $M_n c^2$.) The electronic term $e_e$ is
the same for all geometries, and therefore is as in Ref.~\cite{pea18}. We now
examine in detail the first two terms.

\subsection{Skyrme Energy}

The first term of Eq.~(\ref{1}) can be written as an integral over the cell
of an energy density ${\mathcal E_\mathrm{Sky}}(\pmb{r})$, thus
\beqy\label{2}
e_\mathrm{Sky} = \frac{1}{A}\int_{\mathrm{cell}}
      {\mathcal E_\mathrm{Sky}}(\pmb{r})\,d^3\pmb{r} \quad ;
\eeqy
in the case of cylindrical shapes the integration is taken over unit length, 
and in the case of plates over unit area. For our generalized Skyrme 
functionals the energy density ${\mathcal E_\mathrm{Sky}}(\pmb{r})$ is given by
Eq. (A3) of Ref.~\cite{cgp09} in terms of the number densities $n_q(\pmb{r})$, 
the kinetic-energy densities $\tau_q(\pmb{r})$ and the spin-current densities 
$\pmb{J}_q(\pmb{r})$, where $q=p$ or $q=n$ denotes protons or neutrons, 
respectively. 
Note that all the functionals used in Ref.~\cite{pea18} drop the quadratic 
terms in the spin current (thus removing spurious instabilities~\cite{cg10}), 
along with the Coulomb exchange term for protons~\cite{gcp13}; dropping this 
latter term leads to a significant improvement in the mass fits, especially 
mirror-nucleus differences, and can be interpreted as simulating neglected 
effects such as Coulomb correlations, charge-symmetry breaking of the nuclear 
forces and vacuum polarization~\cite{gp08}. 
The ETF method then approximates $\tau_q(\pmb{r})$ and $\pmb{J}_q(\pmb{r})$
as functions of the number densities $n_q(\pmb{r})$ and their 
derivatives; for a convenient summary of the relevant ETF expressions of 
Refs.~\cite{bgh,bbd} see the Appendix of Ref.~\cite{opp97}.
However, when ${\mathcal E_\mathrm{Sky}}$ is replaced by its ETF
approximation ${\mathcal E}_\mathrm{Sky}^\mathrm{ETF}$ all shell effects are 
lost; the ETFSI method restores them perturbatively, as well as adding a
pairing correction. We then have
\beqy\label{3}
e_\mathrm{Sky}^\mathrm{ETFSI}
= \frac{1}{A}\left\{\int_{\mathrm{cell}}
      {\mathcal E_\mathrm{Sky}^\mathrm{ETF}}(\pmb{r})\,d^3\pmb{r} 
+ E^\mathrm{sc, pair}_p + E_\mathrm{pair}\right\}  \, ,
\quad
\eeqy
in which $E^\mathrm{sc, pair}_p$ is the Strutinsky-integral shell 
correction, as modified by pairing, and $E_\mathrm{pair}$ is the BCS 
energy~\cite{pcpg15}.

To further speed up the computations, we avoid solving the Euler-Lagrange
equations by parametrizing the neutron and proton density distributions 
$n_q(\pmb{r})$. We adopt a simple generalization of the form taken in the 
spherical calculations~\cite{ons08}, with a sum of a constant 
``background'' term and a ``cluster'' term according to
\beqy\label{4}
n_q(\xi) = n_{\mathrm{B}q} + n_{\Lambda q}f_q(\xi)  \quad ,
\eeqy
in which
\beqy\label{5}
f_q(\xi) = \frac{1}{1 + \exp \left[\Big(\frac{C_q - R}
{\xi - R}\Big)^2 - 1\right] \exp \Big(\frac{\xi-C_q}{a_q}\Big) }
\eeqy
and $\xi$ denotes the radial coordinate $r$ in the case of spherical cells, the 
radial coordinate $\eta$ in the case of cylindrical cells and $z$, the 
Cartesian coordinate for plates, assumed to lie in the $x-y$ plane. The 
parameter $R$ likewise represents the radius of the spherical cell, the radius 
of the cylindrical cell or the semi-thickness of the plate-like cell. 
At $\xi=C_q$, $f_q=1/2$, whence the parameter $C_q$ has the meaning of a
characteristic size of a nuclear cluster or ``bubble'', the latter being 
a local depletion of the nucleon density below the background level
$n_{\mathrm{B}q}$, which occurs if  $n_{\Lambda q}<0$. 

Evaluation of the integral appearing in Eq.~(\ref{3}) then proceeds in exactly 
the same way for the pasta phases as for the spherical case, except that the 
expressions for the volume element in the integral, the gradient and the 
Laplacian that occur in the ETF expansion have to be chosen appropriately.

The parametrization (\ref{5}) suffers from the formal defect of a kink at the
origin $\xi = 0$. The actual density distributions, of course, show no such 
discontinuity in their derivatives, but no problem will arise with our
numerical integrations (performed with the Gauss-Legendre method), provided
the mesh size is not too small. We find, in fact, that the computed values 
of the integrals remain stable against a reduction of the mesh size down to
0.01 fm, one hundredth of the nucleon radius; our final computations were made
with a mesh size of 0.1 fm. The integrals thus calculated correspond to the 
kink in the parametrization (\ref{5}) having been smoothed out locally,         
over the region $0 < \xi \lesssim 0.01$ fm.

The argument for neglecting neutron shell corrections in the spherical-cell 
calculations~\cite{pea18} is equally valid here. However, while in the
spherical-cell calculations of Ref.~\cite{pea18} we did calculate the proton
shell corrections, we are not yet able to calculate them for non-spherical cell
geometries. Actually, we argued in Ref.~\cite{pea18} that once proton drip sets
in the unbound proton s.p.{} states will form a quasi-continuum, and proton 
shell effects should largely vanish, exactly as do neutron shell effects at all
densities in the inner crust, i.e., beyond the neutron-drip point. We
therefore adopted in Ref.~\cite{pea18} the prescription of dropping the proton 
shell corrections above the proton drip point, and the pairing corrections 
along with them.

Now even for nonspherical cells the density $\bar{n}$ at the proton drip point
is easily determined.  Classically (neglecting quantum tunneling), for protons 
to be able to  escape their chemical potential $\mu_p$ must be greater or equal
to the proton s.p.{} field at the cell surface
\beqy\label{6}
\mu_p \ge U_p(\xi = R) + M_p c^2  \, .
\eeqy
The former quantity is easily calculated by Eqs. (7) and (8) of 
Ref.~\cite{pea18}, since the necessary condition of beta equilibrium is
satisfied, while the latter is given by Eq. (A11) of Ref.~\cite{cgp09}. (This
characterization of the proton drip point is equivalent to the one that we
adopted in Ref.~\cite{pea18}, but easier to implement.) However, we will
see (next section) that once nonspherical shapes are admitted the 
condition~(\ref{6}) is not satisfied anywhere in the inner crust except perhaps
in a narrow region close to the interface with the homogeneous core, at least 
for the functional BSk24 considered here. 

However, there are grounds for expecting proton shell effects to be small for 
pasta phases, even though the protons may be bound within the WS cell. The 
point is that in the case of spaghetti the motion along the symmetry axis 
is unbound, while for lasagna it is the motion in the $x-y$ plane that is
unbound. In both cases the result is that the s.p.{} proton spectrum is
continuous, thus satisfying the criterion we have already been following for 
neglecting both shell and pairing corrections in the case of neutrons and 
dripped protons. Some support for this
conclusion is found in the recent self-consistent band calculations on
lasagna~\cite{kn19}; presumably the shell effects would be somewhat stronger
for spaghetti, where the unbounded motion is only one-dimensional. Moreover,
these calculations do not include pairing, the effect of which is to dampen
the contribution of shell effects on the total binding energy~\cite{pcpg15}.
Thus to the extent that this argument is correct it becomes meaningful to 
compare the pasta results with the full ETFSI+BCS version of our 
spherical results. 

Another reason to anticipate smallness of the shell corrections in the pasta
phases is that there are unbound protons even though Eq.~(\ref{6}) is not
satisfied. Note that Eq.~(\ref{6}) assumes the classical particle motion, but
the proton wave functions at the high densities, as appropriate to the mantle
layers, can substantially penetrate into the neighboring WS cells because of
quantum tunneling. In our calculations this effect is mimicked by the
background term $n_{\mathrm{B}p}$ in Eq.~(\ref{4}). While $n_{\mathrm{B}p}$ is
negligibly small at low densities close to the neutron drip, this background
term becomes appreciable as $\bar{n}$ approaches $n_\mathrm{cc}$, where
$n_\mathrm{cc}\sim0.08$ fm$^{-3}$ is the number density at the crust-core
transition. This leads to increasing number of ``free protons'' 
$Z_\mathrm{free}\equiv Z-Z_\mathrm{cl}$, where $Z$ is the total number of 
protons in the WS cell and $Z_\mathrm{cl}$ is the number of protons clustered 
near the center.

Nevertheless, the analogy between pasta protons on the one hand and neutrons 
and dripped protons on the other is not exact. Because the protons are still
bound in their cells their motion in the $x-y$ plane in the case of spaghetti 
or along the $z$ axis in the case of lasagna is still discretized. Thus the 
continuous s.p. spectrum actually consist of a superposition of continuous s.p.
spectra, each one based on a different discrete state. As a result, even though
the s.p. spectrum is still continuous, the degeneracy changes discontinuously, 
and weak shell-model fluctuations can be expected.

To summarize the situation, while it is likely that shell and pairing 
corrections are smaller for pasta protons than for protons in the spherical 
configuration, it is far from clear that they will be negligible. Thus in 
addition to comparing the pasta results with the full ETFSI+BCS version of 
our spherical results we shall also make the traditional comparison with the 
ETF version. This is the way in which most previous studies of pasta have been
performed (the exceptions include 
Refs.~\cite{oy94,gm,newt}). In this way we will
acquire some idea of the possible impact of shell and pairing effects in pasta,
although no definite conclusion will be possible before they are actually 
calculated. 

\subsection{Coulomb energy}

For the second term in Eq.~(\ref{1}) we denote by 
$n_\mathrm{ch}(\pmb{r})=n_p(\pmb{r})-n_e$ the globally neutral charge 
distribution of protons and electrons in units of the elementary charge $e$. 
Then, as shown in Appendix~\ref{appA}, we have the 
following expressions for the three different geometries.

Spheres:
\beqy\label{7} 
e_\mathrm{C} = \frac{8\pi^2e^2}{A}\int_0^R\left(\frac{u(r)}{r}\right)^2\,dr \quad,
\eeqy
where
\beqy\label{8}
u(r) = \int_0^r\,n_\mathrm{ch}(r^\prime){r^\prime}^2\,dr^\prime   \quad  .
\eeqy

Cylinders:
\beqy\label{9}
e_\mathrm{C} = \frac{4\pi^2e^2}{A}\int_0^R \frac{u(\eta)^2}{\eta}\,d\eta
 \quad,
\eeqy
where
\beqy\label{10}
u(\eta) = \int_0^\eta\,n_\mathrm{ch}(\eta^\prime) \eta^\prime\,d\eta^\prime   \quad  .
\eeqy

Plates:
\beqy\label{11}
e_\mathrm{C} = \frac{4\pi\,e^2}{A}\int_0^R\, u(z)^2\,dz
 \quad,
\eeqy
where
\beqy\label{12}
u(z) = \int_0^z\,n_\mathrm{ch}(z^\prime)\,dz^\prime   \quad  .
\eeqy

As in all our EoS calculations, a correction for the finite size of the proton
is made, as described in Ref.~\cite{pea91}.

\section{Results}
\label{resu}

With the parametrization defined by Eqs.~(\ref{4}) and~(\ref{5}) there are eight
independent geometric parameters for given density $\bar{n}$, or six if $Z$ and
$A$ are specified as well. Our computational procedure here is as described in 
Ref.~\cite{pea18}: for a suitable range of fixed values of $Z$ we automatically
minimize the total ETF energy per nucleon, 
\beqy\label{13}
e^\mathrm{ETF} = e_\mathrm{Sky}^\mathrm{ETF} + e_\mathrm{C} + e_e
- Y_p\,Q_{n,\beta}    \quad  ,  
\eeqy
with respect to six geometric variables and $A$. The complete results of these
computations will be found in the Supplemental Material. 

For each value of the mean density $\bar{n}$ the optimal value of $Z$ is then 
picked out by inspection, and the corresponding value of $e^\mathrm{ETF}$
shown in columns 2 and 3 of Table~\ref{tab1} for cylindrical and plate shapes, 
respectively (the equilibrium values of $Z$ are shown in Table~\ref{tab3}). 
Reliable pasta solutions could not be found outside the range of densities 
shown. Referring to our complete numerical results presented in the 
Supplemental Material it will be seen that at high densities close to the 
interface with the homogeneous core there are values of $Z$ for which the 
calculated energy is significantly lower than the value we have selected. These
cases, which are easily recognized, are associated with very low values of the 
geometrical parameter $C_q$, which imply very steep density gradients, and thus
a failure of the ETF expansion to converge.

Columns 4  and 5 of Table~\ref{tab1} show respectively the optimum ETF and 
ETFSI+BCS values of the energy per nucleon assuming a spherical configuration
(see the Supplemental Material).
In this table the latter are all lower than the former, i.e., the
shell and pairing corrections are all negative here, but this is not a general 
feature. For $\bar{n}\gtrsim 0.073$ fm$^{-3}$ proton drip for spherical cells
occurs~\cite{pea18}, and we assume in our model that the shell and pairing
corrections vanish~\cite{pea18}. Beyond $\bar{n}$ = 0.0749994 fm$^{-3}$ 
the spherical solutions become mechanically unstable, with increasing mean 
density leading occasionally to reduced pressure. 
Note that below 
proton drip the optimal values of $Z$ are different for ETF and ETFSI+BCS. 

Using the method of Ducoin et al.~\cite{duc07}, we showed in Ref.~\cite{pea18} 
that the transition to a homogeneous solution should occur at $\bar{n}$ = 
0.0807555~fm$^{-3}$ for the BSk24 functional. The fact that we can obtain no 
reliable solution either of pasta or spherical form when 
$\bar{n} >  0.0777$~fm$^{-3}$ suggests that there is a narrow range of 
densities that our codes cannot explore. In any case, the calculations of 
Ref.~\cite{duc07} being based on the Thomas-Fermi approximation considering 
small sinusoidal density fluctuations do not necessarily yield the exact 
transition density.

\begin{table}
\centering
\caption{Energy per nucleon (in MeV) for different cell shapes; sphere(1) 
denotes ETF value for spherical shape, sphere(2) denotes same with 
shell and pairing
 corrections added. 
A corresponding optimal shape
is denoted by s 
(spherical), c (cylindrical), and p (plate-like).}
\label{tab1}
\begin{tabular} {|c|cccc|}
\hline 
$\bar{n}$  & cylinder     & plate & sphere(1) & sphere(2)  \\
\hline
0.0490000  &    7.00597    &    & 7.00575 s & 7.00571 s    \\
0.0500000  &    7.05963    &    & 7.05964 c & 7.05952 s    \\
0.0510000  &    7.11249    &    & 7.11273 c & 7.11254 c    \\
0.0520000  &    7.16462    &    & 7.16506 c & 7.16479 c     \\
0.0540000  &    7.26677    &    & 7.26760 c & 7.26715 c\\
0.0560000  &    7.36638    &    & 7.36756 c & 7.36692 c\\
0.0580000  &    7.46373    &    & 7.46521 c & 7.46436 c   \\
0.0600000  &    7.55907    &    & 7.56081 c & 7.55973 c   \\
0.0610000  &    7.60605    &    & 7.60791 c & 7.60670 c   \\
0.0620000  &    7.65262    &    & 7.65458 c & 7.65323 c    \\
0.0630000  &    7.69880    &    & 7.70084 c & 7.69936 c   \\
0.0640000  &    7.74460    &    & 7.74676 c & 7.74509 c   \\
0.0650000  &    7.79007    &    & 7.79226 c & 7.79044 c   \\
0.0660000  &    7.83521    &    & 7.83745 c & 7.83545 c   \\
0.0670000  &    7.88005    &    & 7.88232 c & 7.88011 c   \\
0.0680624  &    7.92737    &    & 7.92967 c & 7.92720 s   \\
0.0691445  &    7.97527    &    & 7.97756 c & 7.97480 s   \\
0.0692552  &    7.98016    &    & 7.98244 c & 7.97965 s   \\
0.0698092  &    8.00455    &    & 8.00682 c & 8.00385 s   \\
0.0703677  &    8.02907    &    & 8.03131 c & 8.02815 s   \\
0.0709307  &    8.05371    &  8.05474 & 8.05592 c &   8.05255 s \\
0.0714981  &    8.07847    &  8.07919 & 8.08065 c &   8.07703 s \\
0.0720701  &    8.10336    &  8.10379 & 8.10549 c &   8.10237 s \\
0.0726466  &    8.12836    &  8.12853 & 8.13044 c &   8.12945 c \\
0.0732278  &    8.15350    &  8.15342 & 8.15551 p &   8.15551 p\\
0.0738136  &    8.17877    &  8.17846 & 8.18069 p &   8.18069 p\\
0.0744042  &    8.20416    &  8.20367 & 8.20599 p &   8.20598 p\\
0.0749994  &    8.22968    &  8.22902 & 8.23139 p &   8.23139 p\\
0.0755994  &    8.25531    &  8.25454 &         p &           p\\ 
0.0762042  &  8.28107    &  8.28022 &         p &           p\\ 
0.0768139  &    8.30693    &  8.30603 &         p &           p\\
0.0774283  &    8.33289    &  8.33201 &         p&            p\\
0.0777000  &    8.34435    &  8.34349 &         p&            p\\
\hline
\end{tabular}
\end{table}

\subsection{Phase transitions and the equation of state}
\label{sect:EoS}

Comparing columns 2 and 3 of Table~\ref{tab1} with each of columns 4 and 5 
determines the energetically preferred shape at each density, which we indicate
by s (spherical), c (cylindrical) or p (plate). In the pasta phases the ETF 
values of the energy per nucleon differ at most by 0.05\% from the values 
determined for spherical cells, which means that the analytic fit (C1) given in
Ref.~\cite{pea18} will remain valid even when we allow the WS cells to be 
non-spherical. 

We recall that the fit in Ref.~\cite{pea18} was designed to be 
used in a uniform manner throughout the entire neutron star, from its outer 
crust to the core. Then the accuracy within $\sim1$\% of that fit was 
sufficient for this purpose. However, it would not allow us to study the 
\emph{differences} between the energies in the three different phases. For this
purpose we have constructed separate, more accurate fits for each phase, 
applicable in a restricted density range, specifically around the expected
densities of the mantle, 0.05 fm$^{-3}\lesssim\bar{n}<n_\mathrm{cc}$. For the 
energy per baryon $e$, this fit reads
\begin{equation}
 e
 = a_1 + a_2 x + \frac{a_3 x}{(1+a_4 x)^3},
\label{Efit}
\end{equation}
where $x\equiv\bar{n}/n_\mathrm{cc}$ is the natural dimensionless
density argument in the mantle, and the coefficients $a_i$ are listed in
Table~\ref{tab:Efit}.

\begin{table}
\caption{Parameters of Eq.~(\ref{Efit}) for different WS cell shapes; 
sphere(1) and sphere(2)
denote ETF and ETFSI+BCS values for the spherical shape, respectively.}
\label{tab:Efit}
\begin{tabular} {|c|cccc|}
\hline 
cell shape     & $a_1$   & $a_2$  & $a_3$  & $a_4$   \\
               & (MeV)   & (MeV)  & (MeV)  & \\
\hline
sphere(1)      & 1.76708 & 4.1198 & 14.026 & 0.75683 \\
sphere(2)      & 1.42067 & 4.2803 & 16.124 & 0.79924 \\
cylinder       & 1.57416 & 4.2769 & 15.413 & 0.80464 \\
plate          & 2.27349 & 4.1331 & 12.655 & 0.82983 \\
\hline
\end{tabular}
\end{table}

\begin{figure}
\includegraphics[width=\columnwidth]{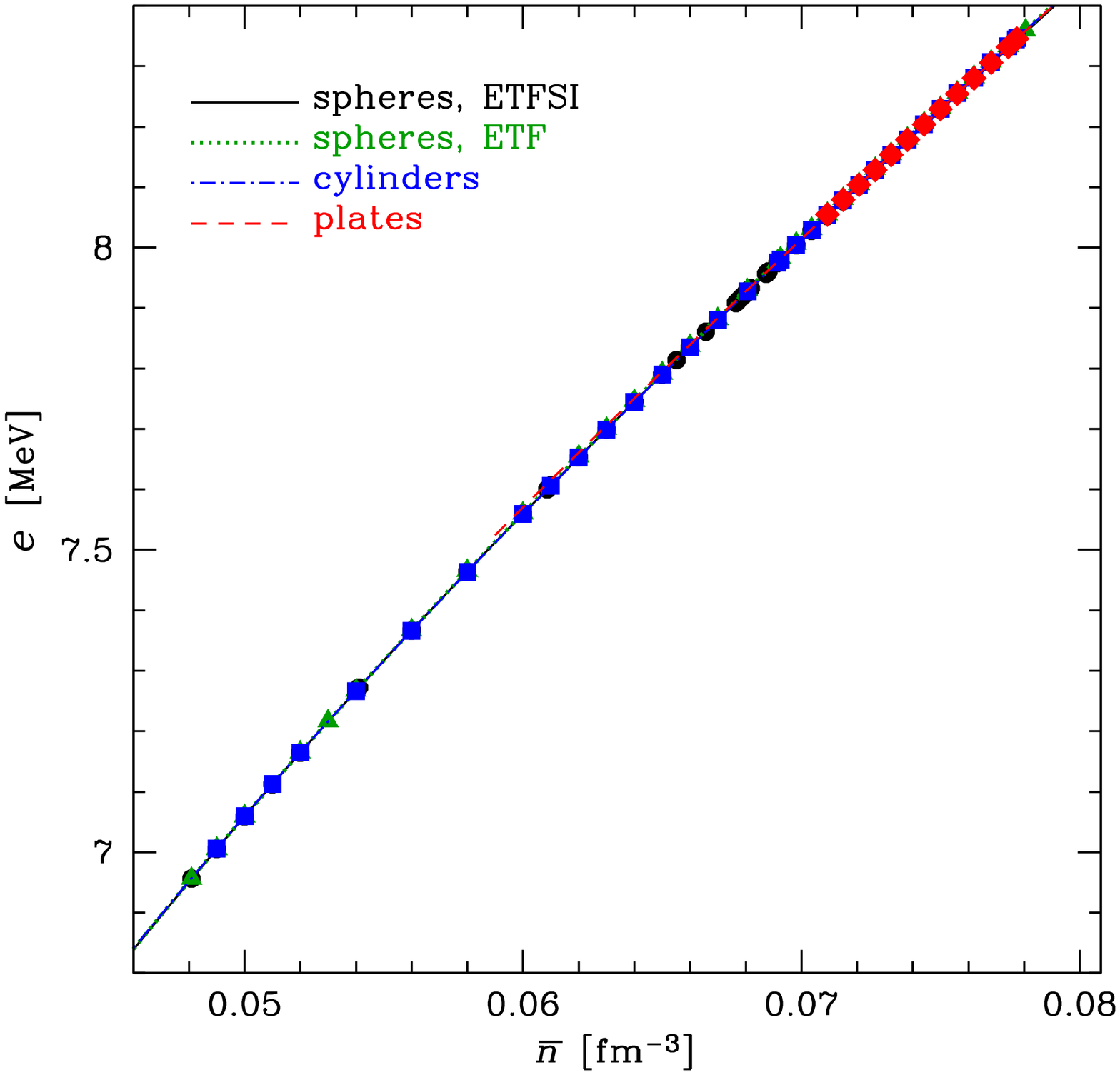}
\caption{Energy per baryon as function of the mean baryon density. Symbols show
the calculated values and lines show the fit (\ref{Efit}) for different WS cell
shapes: spherical with shell and pairing corrections included (black dots and 
solid lines) or excluded (green triangles and dotted lines), cylindrical (blue 
squares and dot-dashed lines), and plate-like (red diamonds and dashed lines).} 
\label{fig:E}
\end{figure}

\begin{figure}
\includegraphics[width=\columnwidth]{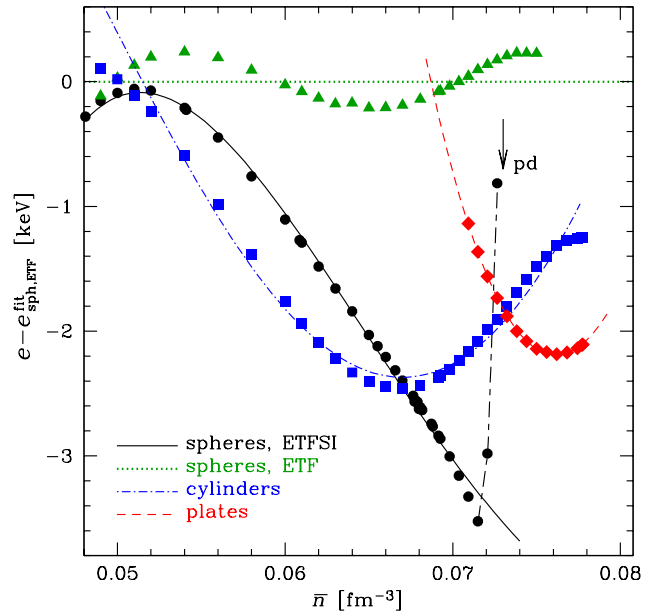}
\caption{The same as in Fig.~\ref{fig:E}, but, instead of energy $e$, the 
difference $e-e_\mathrm{sph,ETF}^\mathrm{fit}$ is shown, where 
$e_\mathrm{sph,ETF}^\mathrm{fit}$ is the fit (\ref{Efit}) for the spherical WS 
cells in the ETF approximation. The long-dash--short-dash line connects the 
ETFSI+BCS results for the spheres at $\bar{n}>0.07207$, where $Z$ changes
 and the numerical results depart from the fit (\ref{Efit}). The 
arrow, marked ``pd,'' points to the density of the proton drip in the spherical
geometry according to Ref.~\cite{pea18}.} 
\label{fig:difE}
\end{figure}

In Fig.~\ref{fig:E} we show the calculated energy per baryon as a function of 
the mean baryon density and the fits. As noted above, the differences between 
the different cell shapes are almost
indistinguishable, which means that the energy density is very weakly sensitive
to the assumed WS cell shape. This makes the phase transitions sensitive to the
small corrections to the energy, depending on the theoretical model (such as
ETF or ETFSI+BCS for the spherical cells, in our case). In order to make a 
choice between the phases, one should consider the differences between the 
energy values. These differences are visualized in Fig.~\ref{fig:difE} by 
showing the energy after subtraction of a common background function, which is 
chosen to be the fit (\ref{Efit}) for the spherical WS cells in the ETF theory 
(the first row of Table~\ref{tab:Efit}). We see that the cylindrical shape 
becomes energetically preferred, that is the ``sphere $\rightarrow$ cylinder''
transition occurs, at $\bar{n}\approx$(0.050\,--\,0.051) fm$^{-3}$. Within the
ETFSI formalism, at higher densities $\bar{n}$ the energy for the spheres
increases less steeply than both for the cylinders and for the spheres in the
ETF formalism, which is revealed in the increasingly steep descent of the 
ETFSI+BCS curve in Fig.~\ref{fig:difE}.  Eventually, at $\bar{n}\approx0.067$ 
fm$^{-3}$ the ETFSI+BCS energy for the spheres again becomes lower than the 
ETF energy for the cylinders, that is, the back-transition ``cylinder 
$\rightarrow$ sphere'' occurs. Note that with the ETFSI+BCS method the number 
of protons per spherical WS cell is constant, $Z=40$, for 
$\bar{n}\lesssim 0.072$ fm$^{-3}$ (see Ref.~\cite{pea18}), whereas the ETF 
method leads to non-integer continuously changing $Z$. The deviation of the 
ETFSI+BCS points beyond $\bar{n}\gtrsim0.072$ fm$^{-3}$ in 
Fig.~\ref{fig:difE} (that are connected by a long-dash--short-dash line as a 
guide to the eye) is a consequence of the fact that $Z$ is starting to change 
discontinuously for the ETFSI+BCS calculations at these densities. As a result,
the second  transition ``sphere $\rightarrow$ cylinder'' occurs at $\bar{n}$ 
between 0.0720701 and 0.0726466 fm$^{-3}$. It precedes the proton drip at 
$\bar{n}\approx0.073$ fm$^{-1}$ \cite{pea18} and the transition ``cylinder 
$\rightarrow$ plate,'' which occurs at nearly the same density. In contrast, 
when we compare the ETF results for the spheres and the cylinders, we do not 
observe the back-transition ``cylinder $\rightarrow$ sphere.''

Thus, for certain densities the preferred shape depends on whether we compare 
the ETF results for the cylinders and plates with the ETF or ETFSI+BCS results 
for the spheres, i.e., whether we exclude or include the shell and pairing
corrections for the spherical configuration. While comparison with the ETF 
results yields the usual sequence of shapes with increasing density, ``sphere 
$\rightarrow$ cylinder $\rightarrow$ plate'', over a certain density range 
comparison with the ETFSI+BCS results (shell and pairing corrections 
included) leads to the more complicated sequence of ``sphere $\rightarrow$ 
cylinder $\rightarrow$ sphere $\rightarrow$ cylinder $\rightarrow$ plate.''
Whether or not this rather unusual feature of a back-transition ``cylinder 
$\rightarrow$ sphere'' would survive the inclusion of the shell and pairing 
corrections to the pasta calculations depends very much on their magnitude: an 
inspection of Table~\ref{tab1} shows they would have to exceed 30\% of the 
corrections to the spherical ETF calculations for the back transition to be 
completely eliminated.  

The density of the initial sphere $\rightarrow$ cylinder transition is also 
seen to be \emph{slightly} sensitive to whether we compare with columns 4 or 5: 
including the shell and pairing corrections for the spherical configuration 
shifts the transition density from $\bar{n} = 0.050$~fm$^{-3}$ to 
$0.051$~fm$^{-3}$. But regardless of which option is chosen there is a 
disagreement with the calculations of Martin and Urban using the same method as
described in Ref.~\cite{mu15}; they find that the sphere $\rightarrow$ cylinder
transition occurs at a density  $\bar{n}\simeq 0.057$~fm$^{-3}$ for BSk24 
(private communication). Besides, they predict that the final cylinder 
$\rightarrow$ plate transition occurs at a lower density, at about 
$0.069$~fm$^{-3}$, compared to our estimate of $0.073$~fm$^{-3}$. 

In neither the spherical nor the non-spherical configurations do we find at    
beta equilibrium, i.e., for the equilibrating value of $Z$, any of the          
``inverted'' solutions that were found in the original liquid-drop              
calculations of Ref.~\cite{rpw83}. In this respect we agree with the ETF        
calculations of Martin and Urban using the same BSk24 functional (private       
communication). 

For cylinders proton drip starts for $\bar{n}$ around $0.077$~fm$^{-3}$, a
density at which plates are energetically favored. And since for plates we 
nowhere find reliable solutions with proton drip, it follows that once we
allow pasta shapes for the WS cells, proton drip can occur nowhere in the inner
crust, except perhaps very close to the interface with the core.  

\begin{table}
\centering
\caption {Pressure (in MeV fm$^{-3}$). Columns 2 and 3 refer to spherical 
cells, without and with shell and pairing 
corrections, respectively. Column 4 refers to the equilibrium cell 
shape: s (spherical), c (cylindrical) and p (plate-like).}
\label{tab2}
\begin{tabular}{|c|ccc|}
\hline
$\bar{n}$ &  $P_\mathrm{sph}(1)$ & $P_\mathrm{sph}(2)$  &  $ P_\mathrm{eq}$ \\
\hline
0.0490000  &    0.1330 & 0.1330 &    0.1330  s  \\
0.0500000  &    0.1364 & 0.1363 &    0.1357 c  0.1363  s\\
0.0510000  &    0.1398 & 0.1398 &      0.1393 c \\
0.0520000  &    0.1433  & 0.1433 &    0.1427        c\\
0.0540000  &    0.1505 &  0.1505 &   0.1499         c\\
0.0560000  &    0.1580 &  0.1581 &   0.1573        c\\
0.0580000  &    0.1658 &  0.1659 &   0.1651       c\\
0.0600000   &   0.1739  & 0.1741 &  0.1733         c\\
0.0610000   &   0.1781  & 0.1783 &   0.1775         c\\
0.0620000  &    0.1824 &  0.1826 &   0.1817       c\\
0.0630000   &   0.1868 &  0.1870 & 0.1861       c\\
0.0640000    &  0.1913 &  0.1915 & 0.1906         c\\
0.0650000   &   0.1958 &  0.1962 & 0.1952      c\\
0.0660000    &  0.2005 &  0.2009 & 0.1999     c\\
0.0670000    &  0.2053 &  0.2057 & 0.2048      c\\
0.0680624    &  0.2105  & 0.2110 &   0.2101 c, 0.2110 s\\
0.0691445   &   0.2158  & 0.2164 &   0.2155 c, 0.2164 s\\
0.0692552 &     0.2164  & 0.2169 &   0.2160 c, 0.2169 s\\
0.0698092  &    0.2192  & 0.2198 &  0.2189 c, 0.2198 s\\
0.0703677   &   0.2220  & 0.2226 &   0.2218 c, 0.2226 s\\
0.0709307   &   0.2249  & 0.2255 &  0.2247 c,  0.2255 s\\
0.0714981    &  0.2279 &  0.2284 &   0.2277 c, 0.2284 s\\
0.0720701   &   0.2309 &  0.2313 & 0.2308 c, 0.2313 s\\
0.0726466    &  0.2338  & 0.2338 &   0.2339         c\\
0.0732278   &   0.2368  & 0.2368 &  0.2356     p\\
0.0738136    &  0.2399  & 0.2399 &   0.2390         p\\
0.0744042    &  0.2429  & 0.2429 &  0.2424         p \\
0.0749994     & 0.2459 & 0.2459 &    0.2458         p\\
0.0755994     &&&              0.2493         p\\
0.0762042        &     &&      0.2528         p\\
0.0768139         &   &&       0.2565         p\\
0.0774283              && &    0.2601         p\\
0.0777000           &&  &      0.2616        p \\
\hline
\end{tabular}
\end{table}

Columns 2 and 3 of Table~\ref{tab2} show respectively the pressure (calculated 
as described in App.~B of Ref.~\cite{pcgd12}) for the optimal ETF and 
ETFSI+BCS  spherical configurations, while column 4 shows the pressure for 
the actual equilibrium shape, spherical (s), cylinder (c) or plate (p), as the 
case may be. In the pasta phases the 
pressure differs at most by 0.6\% from the spherical-cell value, which means 
that the analytic fit (C4) of Ref.~\cite{pea18} is still applicable.
Nevertheless, exactly as for the energy per baryon, we have constructed
separate, more accurate fits for each phase, applicable in the restricted
density range around the expected densities of the mantle. They read 
\begin{equation}
   P = a_1 + a_2 x + \frac{a_3 x^8}{1+a_4 x^{12}},
\label{Pfit}
\end{equation}
where, as before, $x\equiv\bar{n}/n_\mathrm{cc}$, and the coefficients $a_i$ 
are listed in Table~\ref{tab:Pfit}. Whereas the fit presented in 
Ref.~\cite{pea18} uniformly covers the entire neutron star and ensures the 
accuracy within 4\%, the fit (\ref{Pfit}) is applicable only at 0.05 
fm$^{-3} \lesssim \bar{n} \lesssim 0.08$ fm$^{-3}$, but provides an accuracy 
within 0.1\% with respect to the numerical data in Table~\ref{tab2}. 
 
\begin{table}
\caption{Parameters of Eq.~(\ref{Pfit}) for different WS cell shapes; sphere(1)
and sphere(2) denote ETF and ETFSI+BCS values for the spherical shape, 
respectively.}
\label{tab:Pfit}
\begin{tabular} {|c|cccc|}
\hline 
cell shape     & $a_1$      & $a_2$   & $a_3$  & $a_4$ \\
 & (MeV fm$^{-3}$) & (MeV fm$^{-3}$) & (MeV fm$^{-3}$) &\\
\hline
sphere(1)      & $-0.01429 $ & 0.2399  & 0.1058 & 1.41  \\
sphere(2)      & $-0.02128 $ & 0.25153 & 0.0934 & 1.35  \\
cylinder       & $-0.02705$ & 0.26    & 0.0785 & 0.866 \\
plate          & $-0.184  $ & 0.463   & 0      & --    \\
\hline
\end{tabular}
\end{table}

\begin{figure}
\includegraphics[width=\columnwidth]{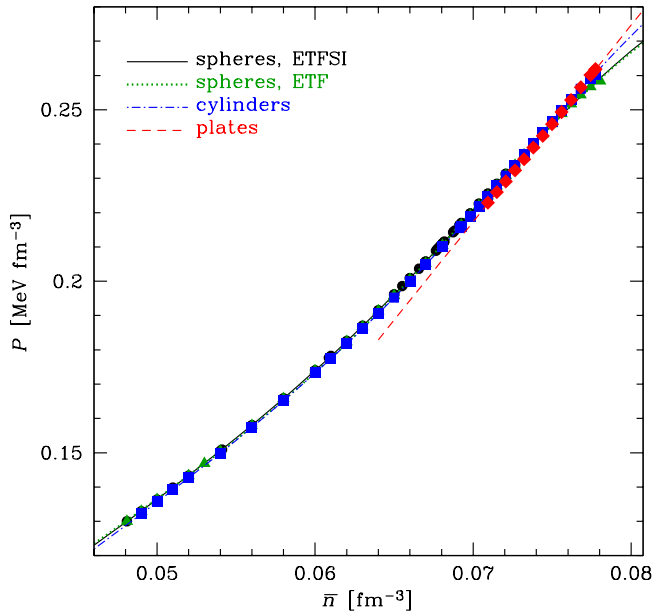}
\caption{Pressure as function of the mean baryon density. Symbols show the 
calculated values and lines show the fit (\ref{Pfit}) for different WS cell 
shapes: spherical with shell and pairing corrections included (black dots and 
solid lines) or excluded (green triangles and dotted lines), cylindrical (blue 
squares and dot-dashed lines), and plate-like (red diamonds and dashed lines).} 
\label{fig:Pfit}
\end{figure}

In Fig.~\ref{fig:Pfit} we show the pressure as function of the mean
baryon density. We see that the pressure values for the different
phases are almost indistinguishable, yet one can discern small differences in
the slope:  the EoS becomes stiffer along the transition sequence ``sphere
$\rightarrow$ cylinder $\rightarrow$ plate.''

\begin{table}
\caption{Parameters of Eq.~(\ref{mufit}) for different WS cell shapes; 
sphere(1) and sphere(2)
denote ETF and ETFSI+BCS values for the spherical shape, respectively.}
\label{tab:mufit}
\begin{tabular} {|c|ccccc|}
\hline 
cell shape     & $a_1$      & $a_2$   & $a_3$  & $a_4$  & $a_5$ \\
               & (MeV)      & (MeV)   & (MeV) &&\\
\hline
sphere(1)      & 5.2813 &  9.2587 & 3.162  & 1.0677 & 0.15509  \\
sphere(2)      & 5.2243 &  9.6969 & 4.358  & 1.0361 & 0.16681  \\
cylinder       & 6.2911 &  6.0102 & 0.8103 & 1.0207 & 0.028241 \\
plate          & 4.9556 & 10.0154 & 3.128  & 1.1295 & 0.20485  \\
\hline
\end{tabular}
\end{table}

\begin{figure}
\includegraphics[width=\columnwidth]{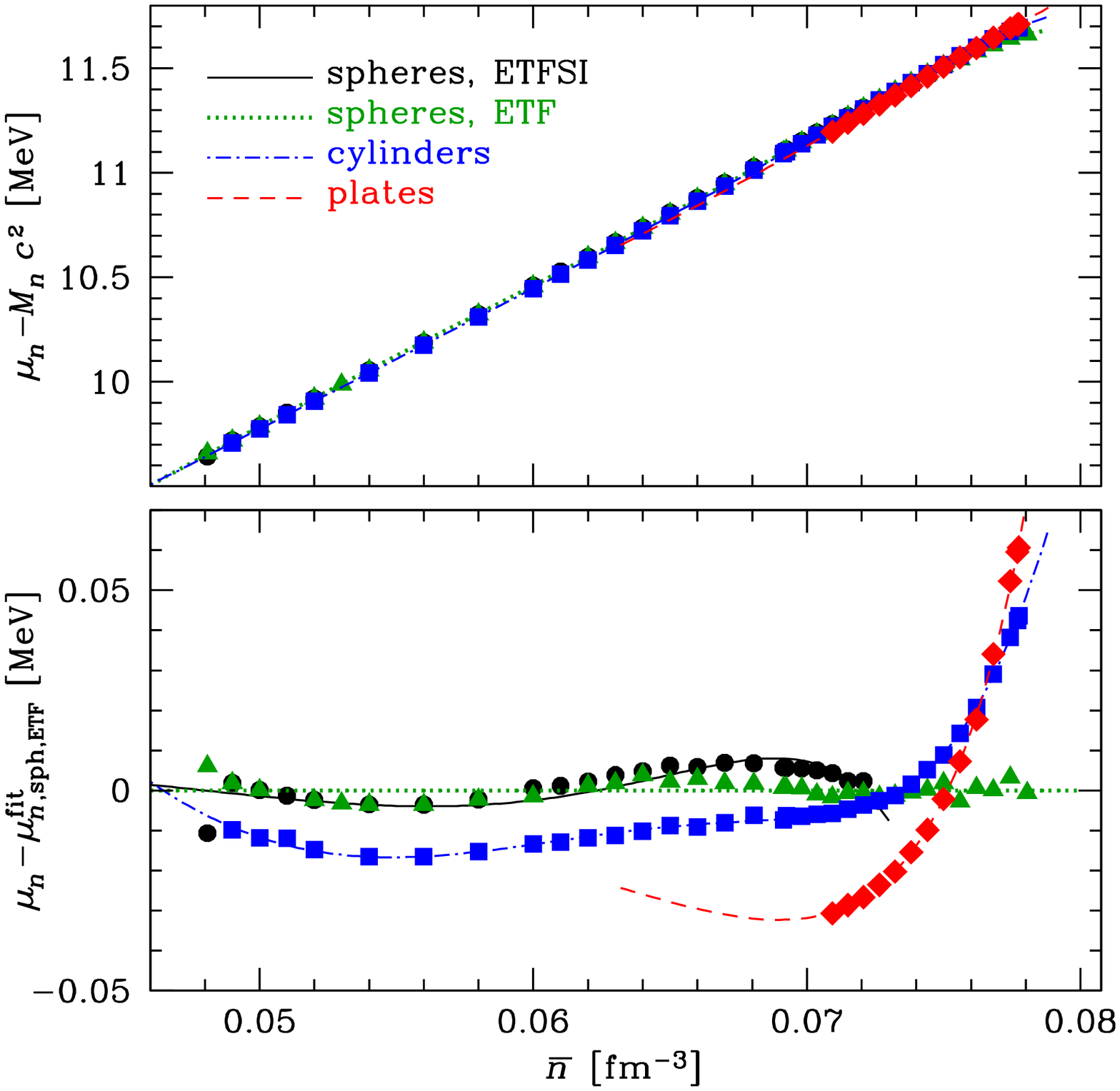}
\caption{Chemical potential of neutrons $\mu_n$ as function of the mean baryon
density. Symbols show the calculated values and lines show the fit (\ref{Pfit})
for different WS cell shapes: spherical with shell and pairing corrections
included (black dots and solid lines) or excluded (green triangles and dotted
lines), cylindrical (blue squares and dot-dashed lines), and plate-like (red
diamonds and dashed lines). \textit{Upper panel}: the difference between
$\mu_n$ and the rest energy of neutron $M_nc^2$.   \textit{Lower panel}: the
difference between $\mu_n$ and the fit (\ref{mufit}) to the ETF results for the
spherical WS cells.} 
\label{fig:mufit}
\end{figure}

In the gravitational field of a neutron star, one needs to know the chemical
potentials of the different particles to determine the chemical equilibrium.
The chemical potential of the strongly degenerate electrons $\mu_e$ is
determined by the formulas given in App.~B of Ref.~\cite{pea18}. The chemical
potential of protons is related to $\mu_e$ and to the chemical potential of
neutrons $\mu_n$ by the condition of beta-equilibrium, $\mu_p=\mu_n-\mu_e$, 
where $\mu_n = g$, the Gibbs free energy per nucleon. For
the chemical potential of neutrons, the fit (C18) of Ref.~\cite{pea18},
originally constructed for the crust, holds also in the mantle, because the
differences between $\mu_n$ in the different phases lie within its accuracy
level. However, for a study of phase equilibrium one needs to know the
differences between $\mu_n$ in the three phases. For this purpose we have
constructed the following analytical approximations:
\begin{equation}
   \mu_n - M_n c^2 = (a_1+a_2 x + a_3 x^8)(a_4-x)^{a_5},
\label{mufit}
\end{equation}
where $x\equiv\bar{n}/n_\mathrm{cc}$, and the coefficients $a_i$ are listed in
Table~\ref{tab:mufit}. As compared to the fit for $\mu_n$ given by 
Eq.~(C18) in Ref.~\cite{pea18}, Eq.~(\ref{mufit}) has a narrow applicability 
range, 0.05 fm$^{-3}\lesssim \bar{n}\lesssim0.08$ fm$^{-3}$ instead of 
$2\times10^{-4}$ fm$^{-3}\lesssim \bar{n}\lesssim0.08$ fm$^{-3}$,  but in 
return it provides an order-of-magnitude better accuracy. The comparison of the
calculated and fitted $\mu_n$ is shown in Fig.~\ref{fig:mufit}. In the upper 
panel the differences between different results are barely
distinguishable. In order to visualize them, in the lower panel we
subtract from each $\mu_n$ value the respective value given  by the fitting
formula (\ref{mufit}) for the spherical WS cells in the ETF approximation (the
first row of parameters in Table~\ref{tab:mufit}). It is seen that in the
vicinity of the transition densities the differences between the respective
values of $\mu_n$ are of the order of 10\,--\,20 keV.

\begin{table}
\centering
\caption {Proton number $Z$ and fraction $Y_p$. Column 2 shows $Z$ value for 
the actual equilibrium shape (note that units depend on shape). Columns 3 and 4 
show $Y_p$ for spherical cells, without and with shell and pairing corrections, 
respectively. Column 4 shows $Y_p$ for the equilibrium cell shape: s 
(spherical), c (cylindrical) and p (plate-like).}
\label{tab3}
\begin{tabular}{|c|cccc|}
\hline
$\bar{n}$ & $Z_\mathrm{eq}$ & $Y_p^\mathrm{sph}(1)$ & $Y_p^\mathrm{sph}(2)$ & $Y_p^\mathrm{eq}$ \\
\hline
0.0490000  &  40  s          & 0.03366   & 0.03358     &     0.03358      s\\
0.0500000  &  1.503\,c, 40\,s   & 0.03354  &  0.03346    & 0.03372\,c, 0.03346\,s\\
0.0510000  &  1.501 c        & 0.03344   & 0.03335   &     0.03369 c  \\
0.0520000  &  1.490 c&    0.03333&  0.03325   & 0.03351         c\\
0.0540000  &  1.482 c&   0.03317 & 0.03308   &  0.03333        c\\
0.0560000   & 1.476 c&   0.03301 & 0.03292  &   0.03318         c\\
0.0580000  &  1.472 c&   0.03289 & 0.03280    & 0.03306         c\\
0.0600000   & 1.472 c&    0.03280&  0.03270 &   0.03296         c\\
0.0610000   & 1.474 c&   0.03277 & 0.03265  &   0.03292    c\\
0.0620000   & 1.476 c&  0.03273&  0.03262     & 0.03289       c\\
0.0630000   & 1.480 c&   0.03272 & 0.03259    & 0.03287         c\\
0.0640000   & 1.484 c&  0.03267 & 0.03257     & 0.03285      c \\
0.0650000   & 1.492 c&  0.03270 & 0.03256   &   0.03284        c \\ 
0.0660000   & 1.528 c&  0.03269 & 0.03255     & 0.03286      c \\
0.0670000  &  1.512 c&  0.03271 & 0.03255  &    0.03284      c  \\
0.0680624 &   1.528\,c, 40\,s&  0.03271 & 0.03256 & 0.03297\,c, 0.03256\,s  \\
0.0691445  &  1.542\,c, 40\,s & 0.03274 & 0.03257 & 0.03287\,c, 0.03257\,s \\
0.0692552  &  1.546\,c, 40\,s&  0.03274 & 0.03258 & 0.03287\,c, 0.03258\,s\\ 
0.0698092   & 1.554\,c, 40\,s&  0.03275 & 0.03259 & 0.03288\,c, 0.03259\,s\\  
0.0703677  &  1.566\,c, 40\,s&  0.03278 & 0.03260 & 0.03290\,c, 0.03260\,s\\
0.0709307   & 1.576\,c, 40\,s&  0.03279 & 0.03262 & 0.03291\,c, 0.03262\,s\\
0.0714981   & 1.590\,c, 40\,s&  0.03281 & 0.03264 & 0.03293\,c, 0.03264\,s  \\
0.0720701 &   1.604\,c, 41\,s&  0.03282 & 0.03267 & 0.03295\,c, 0.03267\,s \\
0.0726466  &  1.618 c &   0.03285 & 0.03296 &  0.03297       c  \\
0.0732278   & 0.0643 p &  0.03288 & 0.03288  & 0.03316        p\\
0.0738136 &   0.0643 p&   0.03290&  0.03290  & 0.03318         p\\
0.0744042  &  0.0643 p &  0.03293 & 0.03294 &  0.03320           p\\
0.0749994  &  0.0646 p&  0.03296 &0.03296& 0.03324         p\\
0.0755994   & 0.0642 p&           &      &  0.03337      p\\
0.0762042   & 0.0649 p&           &      &  0.03342     p\\
0.0768138&    0.0652 p&           &      &  0.03333    p\\
0.0774283   & 0.0657 p&           &      &  0.03337        p\\
0.0777000 &   0.0673 p&           &      &  0.03342         p\\
\hline
\end{tabular}
\end{table}

\subsection{Free and bound neutron and proton numbers}

\begin{figure}
\includegraphics[width=\columnwidth]{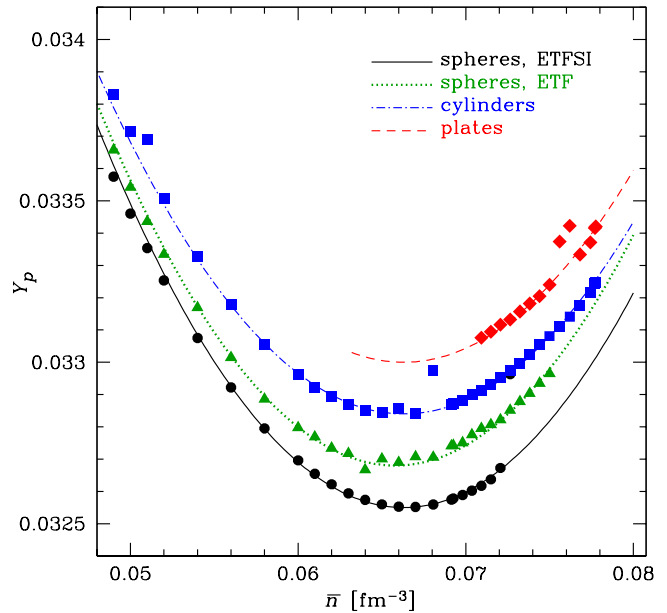}
\caption{Proton fraction as function of the mean baryon density.
Symbols show the calculated values and lines show the fit (\ref{Yfit})
for different WS cell shapes: spherical with shell and pairing corrections
included (black dots and solid lines) or excluded (green triangles and dotted
lines), cylindrical (blue squares and dot-dashed lines), and plate-like (red
diamonds and dashed lines).} 
\label{fig:Yfit}
\end{figure}

Column 2 of Table~\ref{tab3} shows the equilibrium value of $Z$, regardless of 
shape. Since this quantity even has different dimensions for the different 
shapes (for spheres it is the number of protons in the cell, for cylinders the 
number per fm and for plates the number per fm$^2$) there can be no
comparison of the pasta values with the spherical value. On the other hand, 
such a comparison is meaningful for the proton fraction $Y_p = Z/A$, so in 
columns 3 and 4 we display the values of $Y_p$ assuming spherical cells, the
first without shell and pairing corrections and the second with them. In column
5 we show the values of $Y_p$ for the equilibrium pasta shape. Remarkably, 
despite the drastic difference in geometries almost the same values are 
obtained, the difference never exceeding 1\%, which means that we can still use
the analytic fit (C6) of Ref.~\cite{pea18}. Nevertheless,  we have constructed
separate, more accurate fits for each phase, applicable in the restricted
density range, specifically around the expected densities of the mantle.
It turns out that in the considered density range $Y_p$ is well reproduced by
the simple parabola
\begin{equation}
   Y_p = Y_\mathrm{min} +a (\bar{n} - n_\mathrm{min})^2.
\label{Yfit}
\end{equation}
Here, $\bar{n}$ is in units of fm$^{-3}$ and the parameters are listed in
Table~\ref{tab:Yfit}. The comparison of calculated and fitted proton fractions
is shown in Fig.~\ref{fig:Yfit}.

\begin{table}
\caption{Parameters of Eq.~(\ref{Yfit}) for different WS cell shapes; 
sphere(1) and sphere(2)
denote ETF and ETFSI+BCS values for the spherical shape, respectively.
}
\label{tab:Yfit}
\begin{tabular} {|c|ccc|}
\hline 
cell shape & $Y_\mathrm{min}$ &$a$ (fm$^6$) & $n_\mathrm{min}$ (fm$^{-3}$) \\
\hline
sphere(1)      & 0.03267 & 3.56 & 0.0660 \\
sphere(2)      & 0.03255 & 3.54 & 0.0663 \\
cylinder       & 0.03284 & 3.17 & 0.0663 \\
plate          & 0.03301 & 3.17 & 0.0663 \\
\hline
\end{tabular}
\end{table}

In practice, for modeling physical processes in the neutron-star crust and 
mantle and determining their physical properties, one needs to know not only 
$Y_p$, but also the numbers of free and bound neutrons and protons in a WS 
cell. For this purpose we have constructed appropriate fitting formulas.
For the spherical WS cells such fits are given in section C3.1 of 
Ref.~\cite{pea18}; they remain unchanged. For the total number of protons in 
the cylindrical WS cells we have, with an accuracy of a few percent,
\begin{equation}
   Z = (1.835 - 0.554x + 0.732 x^9)\mbox{~fm}^{-1},
\end{equation}
where $x=\bar{n}/n_\mathrm{cc}$. For plate-like cells, at 
$\bar{n}<0.074$ fm$^{-3}$ the proton number is constant, $Z\approx Z_0$,
where $Z_0=0.0643$ fm$^{-2}$. At $\bar{n}>0.074$ fm$^{-3}$, $Z$ increases 
approximately as $Z/Z_0\approx 1+ (11\Delta\bar{n})^4, $
where $\Delta\bar{n}=\bar{n}/\mbox{fm}^{-3}-0.074$.
The number of neutrons $N$ is determined by the identity
\begin{equation}
N = Z\left(\frac{1}{Y_\mathrm{p}} - 1\right) .
\end{equation}

\begin{table}
\caption{Parameters of Eq.~(\ref{Zfree}) for the cylindrical
and plate-like WS cells. }
\label{tab:Zfree}
\begin{tabular} {|c|ccc|}
\hline 
cell shape     & $a_1$   &$a_2$ & $a_3$ \\
\hline
cylinder       & 0.62144 & 1.0133 & 1.2708 \\
plate          & 0.26563 & 1.1712 & 4.5795 \\
\hline
\end{tabular}
\end{table}

\begin{table}
\caption{Parameters of Eq.~(\ref{Ynf}) for the cylindrical and plate-like WS 
cells.}
\label{tab:Ynf}
\begin{tabular} {|c|cccc|}
\hline 
cell shape     & $a_1$   &$a_2$   & $a_3$  & $a_4$    \\
\hline
cylinder       & 0.74483 & 0.0959 & 0.0817 & 26 \\
plate          & 0.77675 & 0.0455 & 0.0446 & 21 \\
\hline
\end{tabular}
\end{table}

\begin{figure}
\includegraphics[width=\columnwidth]{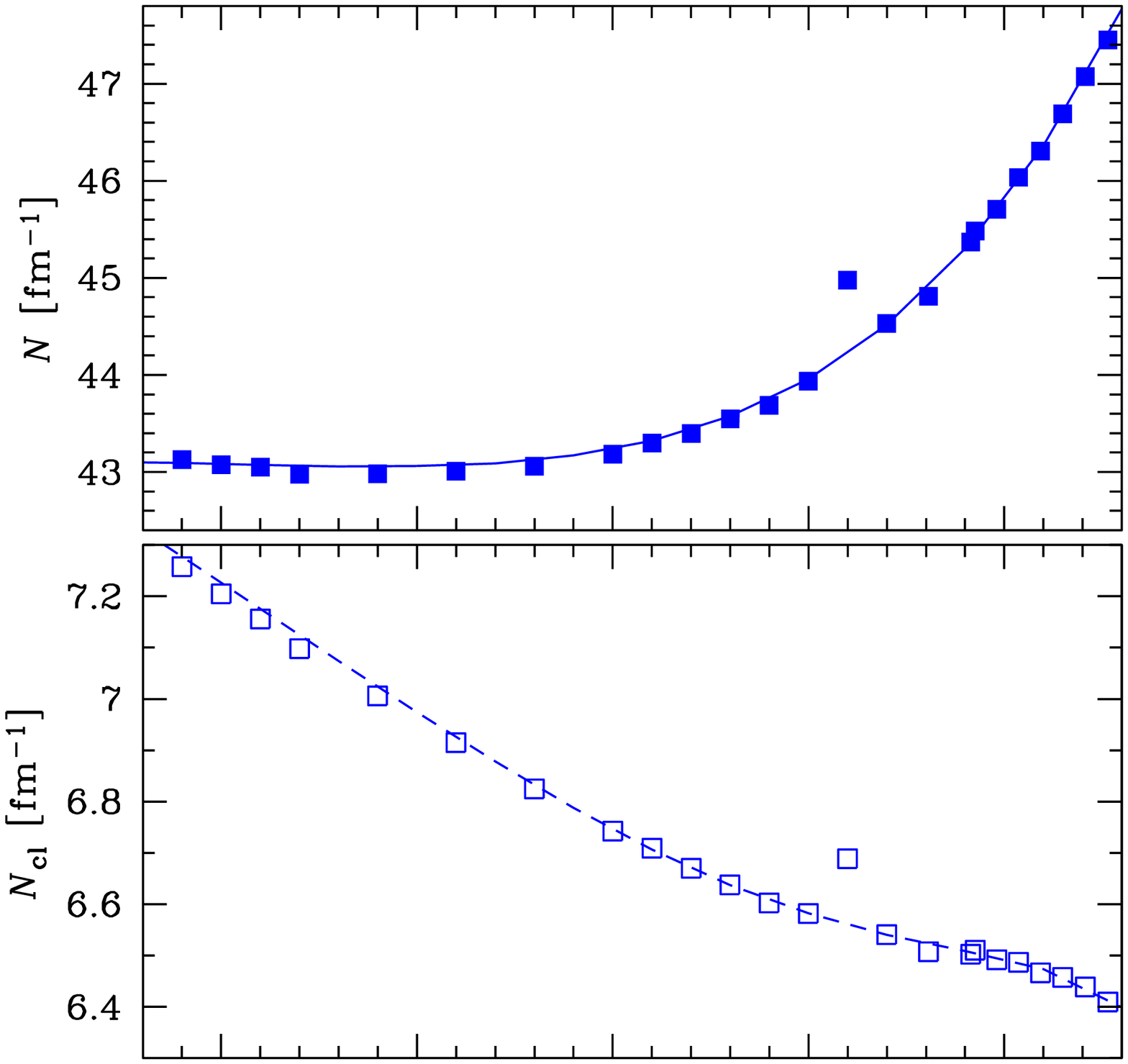}
\includegraphics[width=\columnwidth]{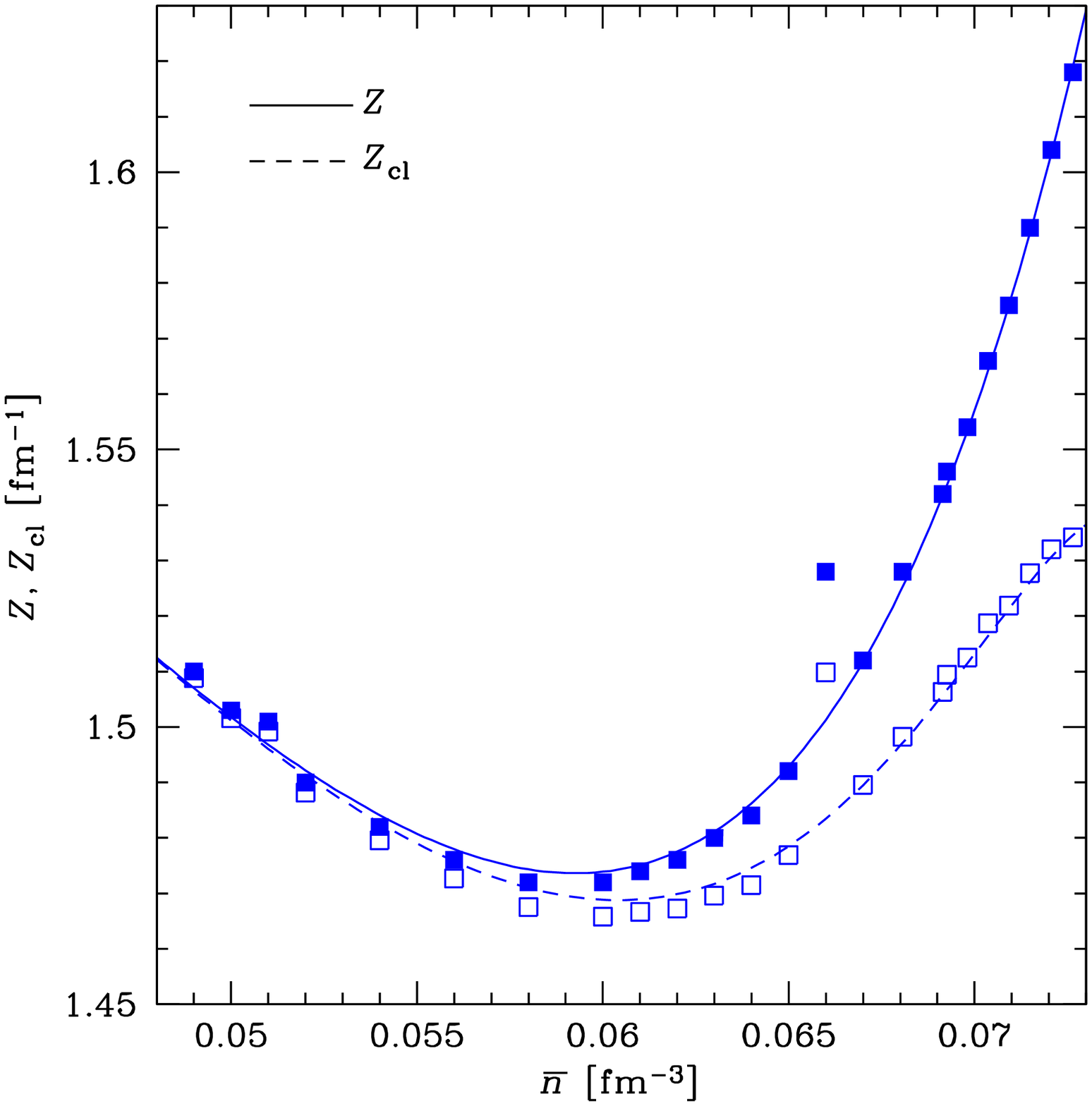}
\caption{Proton and neutron numbers per unit length as functions of mean baryon
density of a cylindrical WS cell.
\textit{Top panel}: all neutrons; \textit{middle panel}: clustered neutrons;
\textit{bottom panel}: all protons (filled symbols and solid lines)
and clustered protons (empty symbols and dashed lines).
The symbols show the calculated values and the lines show the fits.} 
\label{fig:Z}
\end{figure}
\begin{figure}
\includegraphics[width=\columnwidth]{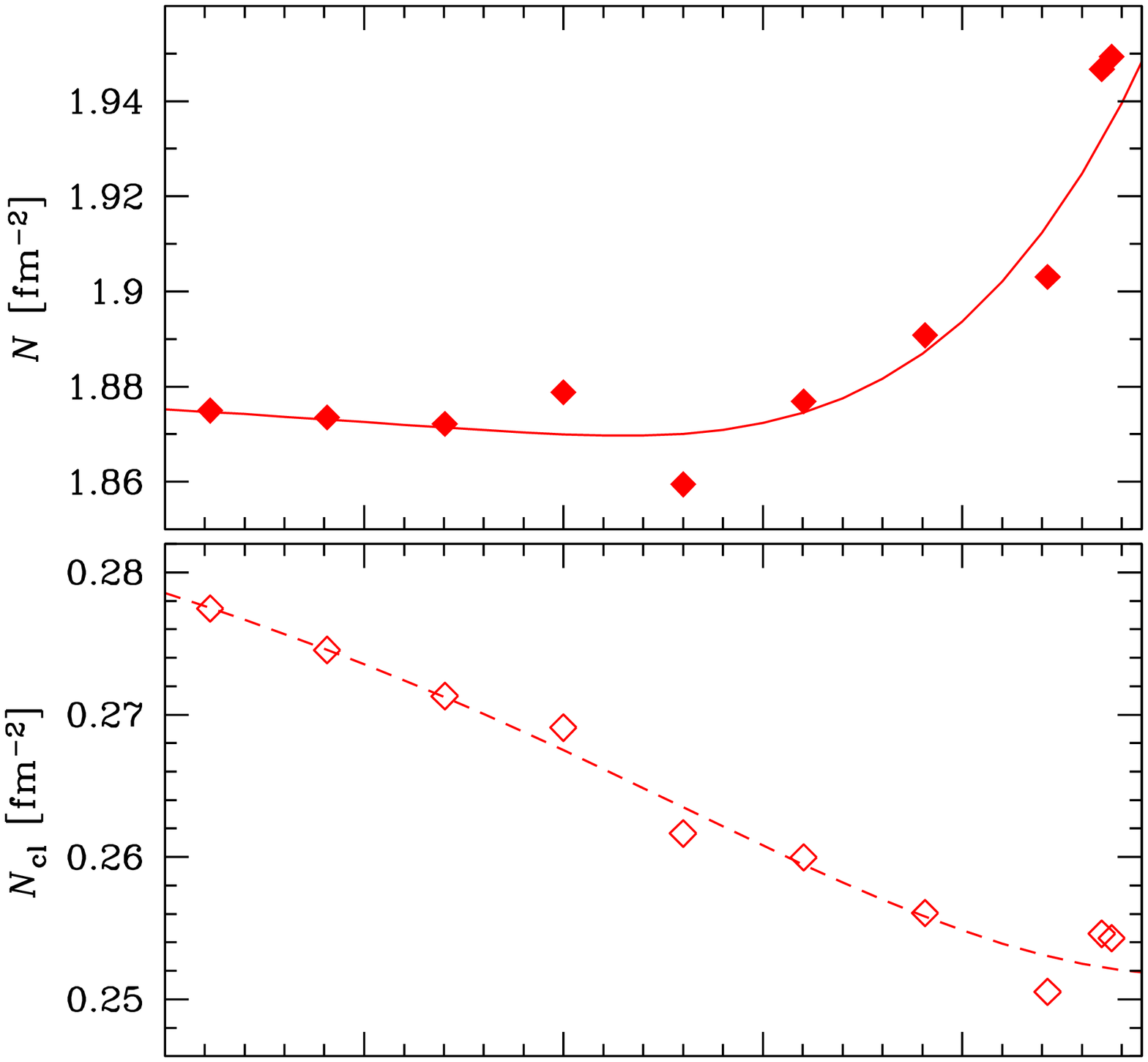}
\includegraphics[width=\columnwidth]{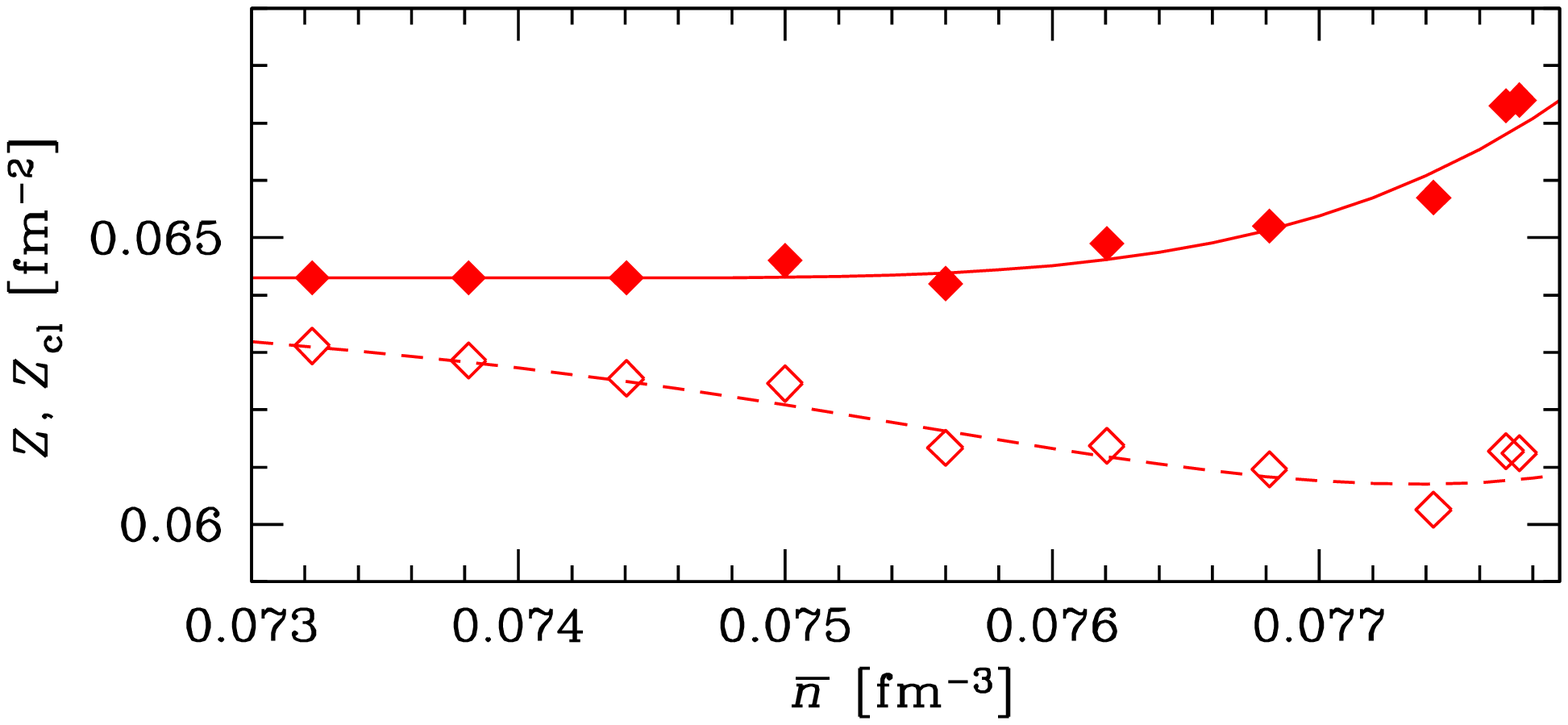}
\caption{Proton and neutron numbers per unit area as functions of mean baryon
density of a plate-like WS cell.
\textit{Top panel}: all neutrons; \textit{middle panel}: clustered neutrons;
\textit{bottom panel}: all protons (filled symbols and solid lines)
and clustered protons (empty symbols and dashed lines).
The symbols show the calculated values and the lines show the fits.} 
\label{fig:N}
\end{figure}

The numbers of unbound neutrons and protons are defined as
$N_\mathrm{free}=n_{\mathrm{B}n}V_\mathrm{cell}$ and
$Z_\mathrm{free}=n_{\mathrm{B}p}V_\mathrm{cell}$,
where $V_\mathrm{cell}=4\pi R^3/3$, $\pi R^2$, and $2R$ for the spheres,
cylinders, and plates, respectively (accordingly, the numbers of free nucleons
are counted per unit length for the cylinders and per unit area for the
plates). The numbers of the neutrons and protons that are bound in clusters are
$N_\mathrm{cl}=N-N_\mathrm{free}$ and $Z_\mathrm{cl}=Z-Z_\mathrm{free}$. The 
numbers of free protons in the cylindrical and plate-like WS cells are 
described by the fit
\begin{equation}
   Z_\mathrm{free}=\frac{(a_1 x)^9}{(a_2-x)^{a_3}},
\label{Zfree}
\end{equation}
where $Z_\mathrm{free}$ is in fm$^{-1}$ and fm$^{-2}$ for the cylinders and
plates, respectively, $x\equiv \bar{n}/n_\mathrm{cc}$, and the parameters $a_i$
are given in Table~\ref{tab:Zfree}. The fraction of free neutrons among all
nucleons $Y_{nf} \equiv N_\mathrm{free}/A$ is approximated as
\begin{equation}
   Y_{nf}=a_1+a_2 x +a_3 x^{a_4},
   \quad
   x\equiv \bar{n}/n_\mathrm{cc},
\label{Ynf}
\end{equation}
with parameters given in Table~\ref{tab:Ynf}.
Now with $Z$ and $Y_p$ already parametrized, the number of free neutrons
is given by the identity $N_\mathrm{free}=Z Y_{nf}/Y_p$.

The calculated and fitted total and bound proton numbers are shown in 
Figs.~\ref{fig:Z} and \ref{fig:N}. Note that although the proton-drip 
threshold, as defined by Eq.(\ref{6}), is not reached in any layer of the 
crust, $Z_\mathrm{free}$ becomes appreciable at densities  $\bar{n} \gtrsim 
0.07$ fm$^{-3}$.
 
\subsection{Nuclear size and shape parameters}

\begin{table}
\caption{Parameters of Eq.~(\ref{Xq}) for the cylindrical and plate-like WS 
cells. Parameters $a_1$ and $a_2$ are in units of fm, $a_3$ is dimensionless. }
\label{tab:Xq}
\begin{tabular} {|c|ccc|ccc|}
\hline 
shape   & \multicolumn{3}{|c|}{cylinder}  & \multicolumn{3}{c|}{plate} \\
parameter  & $a_1$ &$a_2$ & $a_3$ & $a_1$  &$a_2$  & $a_3$  \\
\hline                                                                  
$C_p$      & 5.35  & 5.24 & 6.83  & 4.191  & 4.30  & 16 \\
$C_n$      & 5.97  & 4.73 & 6.53  & 4.799  & 3.82  & 16 \\
$a_p$      & 0.729 & 1.55 & 6.12  & 0.7553 & 0.793 & 10 \\
$a_n$      & 1.214 & 1.37 & 5.09  & 0.9027 & 0.578 & 10 \\
\hline
\end{tabular}
\end{table}

\begin{figure*}
\includegraphics[width=\columnwidth]{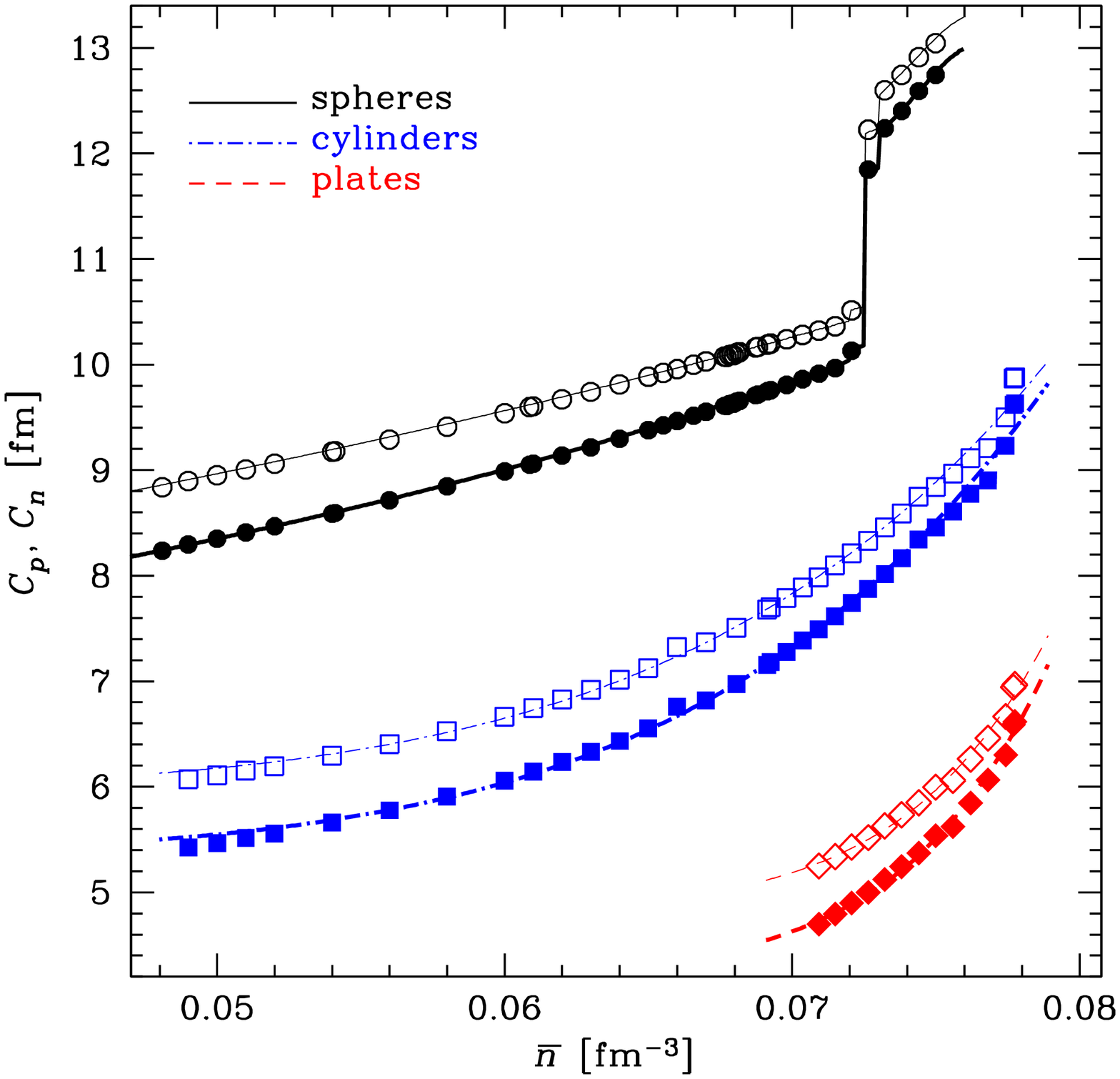}
\includegraphics[width=\columnwidth]{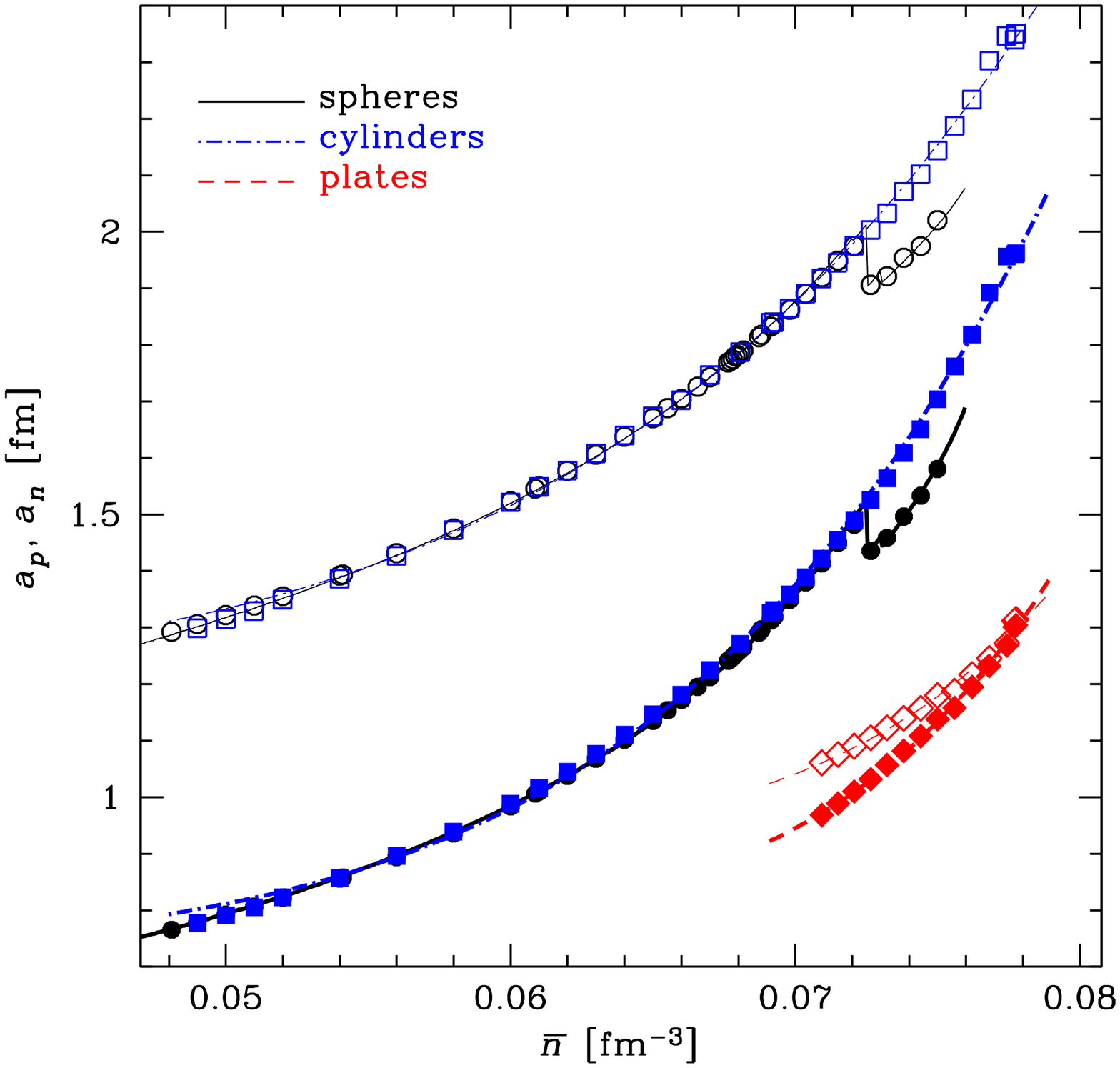}
\caption{Proton (filled symbols) and neutron (empty symbols) size parameters 
$C_q$ (left panel) and diffuseness parameters $a_q$ (right panel) compared to 
the fits (\ref{Xq}) (lines) as functions of mean baryon density for different 
WS cell shapes: spherical (black dots and solid lines), cylindrical (blue 
squares and dot-dashed lines), and plate-like (red diamonds and dashed lines).} 
\label{fig:Xq}
\end{figure*}

For studies of some physical phenomena in the neutron star interiors, it may be
of interest to know the sizes and shapes of the nuclear clusters (see, e.g., 
Ref.~\cite{GYP} for the case of electron heat and charge transport). For this 
purpose, we have constructed analytical approximations to the parameters $C_q$ 
and $a_q$, which determine respectively the size and the diffuseness of a 
cluster, when its density profile is parametrized by Eqs.~(\ref{4}) and 
(\ref{5}). For the spherical WS cells, the fits to $C_q$ and $a_q$ as functions
 of mean baryon density $\bar{n}$ have been published in App.~C5 of
Ref.~\cite{pea18}; they remain unchanged. For the cylindrical and plate-like WS
cells, in the restricted ranges of $\bar{n}$ under consideration, they are
approximated by the simple formula
\begin{equation}
  X_q = a_1+a_2 x^{a_3},
   \quad
   X_q = C_p, C_n, a_p, a_n;
   \quad
   x\equiv \bar{n}/n_\mathrm{cc}.
\label{Xq}
\end{equation}
The fit parameters $a_i$ are given in Table~\ref{tab:Xq}. The calculated and 
fitted $C_p$, $C_n$, $a_p$, and $a_n$
are plotted against $\bar{n}$ in Fig.~\ref{fig:Xq}.

Note that we do not need separate fits to the parameters $n_{\mathrm{B}q}$
and $n_{\Lambda q}$ in Eq.~(\ref{5}), because they are determined by 
the already fitted parameters through the relations
\begin{eqnarray}
   n_{\mathrm{B}p} = \frac{Z_\mathrm{free}}{V_\mathrm{cell}},
&&
   n_{\mathrm{B}n} = \frac{N_\mathrm{free}}{V_\mathrm{cell}},
\\
   n_{\Lambda p} = \frac{Z_\mathrm{cl}}{
                   \int_0^R f_p(\xi) \frac{dV}{d\xi} d\xi},
&&
   n_{\Lambda n} = \frac{N_\mathrm{cl}}{
                   \int_0^R f_n(\xi) \frac{dV}{d\xi} d\xi},
\hspace*{2em}
\\
\mbox{where~}
   \frac{dV}{d\xi} &=& \left\{ \begin{array}{ll} 
    4\pi r^2  & \mbox{(spheres)},   \\
    2\pi \eta & \mbox{(cylinders)}, \\
    2         & \mbox{(plates)}.    \\
    \end{array}
    \right.
\nonumber
\end{eqnarray}

\begin{table*}
\centering
\caption {Inhomogeneities $\Lambda$ and $\Lambda_p$, defined in 
Eqs.~(\ref{14}) and (\ref{15}), respectively.  Columns 2 and 3
show $\Lambda$ for spherical cells, without and with shell and pairing corrections,
respectively; columns 5 and 6 the corresponding values of $\Lambda_p$. Columns 
4 and 7 show $\Lambda$ and $\Lambda_p$, respectively, for the actual 
equilibrium shape.}
\label{tab4}
\begin{tabular}{|c|cccccc|}
\hline
$\bar{n}$ & $\Lambda^\mathrm{sph}(1)$ & $\Lambda^\mathrm{sph}(2)$ &$\Lambda^\mathrm{eq}$ & 
$\Lambda_p^\mathrm{sph}(1)$ & $\Lambda_p^\mathrm{sph}(2)$ &$\Lambda_p^\mathrm{eq}$ \\
\hline
0.0490000 & 0.170 & 0.169   & 0.169 s   & 6.39  & 6.45     & 6.45    s\\
0.0500000 & 0.162 & 0.161   & 0.162 c 0.161 s  & 6.11 & 6.17 & 5.92 c 6.17  s\\
0.0510000 & 0.154 & 0.153   & 0.154 c   & 5.84  & 5.90      & 5.66 c \\
0.0520000 & 0.146 & 0.145 & 0.146 c     & 5.58 & 5.63 & 5.41 c\\
0.0540000 & 0.131 & 0.130 & 0.132 c     & 5.07 & 5.13 & 4.93 c \\
0.0560000 & 0.118 & 0.117 & 0.118 c     & 4.60 & 4.66 & 4.48 c\\ 
0.0580000 & 0.105 & 0.105 & 0.106 c     & 4.16 & 4.22 & 4.05 c\\
0.0600000 & 0.0939& 0.0932& 0.0942 c    & 3.74 & 3.80 & 3.65 c\\ 
0.0610000 & 0.0885& 0.0878 &0.0888 c    & 3.53 & 3.60 & 3.45 c\\
0.0620000 & 0.0833& 0.0825 &0.0836 c    & 3.34 & 3.41 & 3.27 c  \\
0.0630000 & 0.0782 & 0.0775 & 0.0785 c  & 3.14 & 3.22 & 3.08 c  \\
0.0640000 & 0.0731 & 0.0726 & 0.0737 c  & 2.98 & 3.03 & 2.90 c \\
0.0650000 & 0.0686 & 0.0678 & 0.0690 c  & 2.76 & 2.85 & 2.73 c \\
0.0660000 & 0.0639 & 0.0632 & 0.0645 c  & 2.59 & 2.67 & 2.55 c  \\
0.0670000 & 0.0595 & 0.0587 & 0.0600 c          & 2.41 & 2.50 & 2.39 c\\
0.0680624 & 0.0547 & 0.0540 & 0.0554 c, 0.0540 s & 2.23 & 2.31 & 2.20 c, 2.31 s \\
0.0691445 & 0.0501 & 0.0493 & 0.0508 c, 0.0493 s & 2.03 & 2.13 & 2.03 c, 2.13 s  \\
0.0692552 & 0.0496 & 0.0489 & 0.0504 c, 0.0489 s & 2.01 & 2.11 & 2.01 c, 2.11 s   \\
0.0698092 & 0.0472 & 0.0465 & 0.0481 c, 0.0465 s & 1.92 & 2.01 & 1.93 c, 2.01 s  \\
0.0703677 & 0.0449 & 0.0441 & 0.0458 c, 0.0441 s & 1.82 & 1.91 & 1.84 c, 1.91 s  \\
0.0709307 & 0.0425 & 0.0417 & 0.0435 c, 0.0417 s & 1.73 & 1.82 & 1.75 c,1.82 s  \\
0.0714981 & 0.0401 & 0.0392 & 0.0412 c, 0.0392 s & 1.64 & 1.72 & 1.66 c, 1.72 s  \\
0.0720701 & 0.0377 & 0.0369 & 0.0389 c, 0.0369 s & 1.54 & 1.61 & 1.57 c, 1.61 s  \\
0.0726466 & 0.0353 & 0.0353 & 0.0366 c & 1.44 & 1.44 & 1.48 c \\
0.0732278   & 0.0328 &  0.0328 & 0.0327 p   & 1.34 & 1.34 & 1.16 p  \\
0.0738136  &  0.0303 & 0.0303  & 0.0306 p   & 1.24 & 1.24 & 1.09 p  \\
0.0744042   & 0.0277 & 0.0277  & 0.0285 p   & 1.14 & 1.14 & 1.02 p     \\
0.0749994  &  0.0250 & 0.0250 & 0.0265  p     &  1.03 &  1.03&  0.950  p\\
0.0755994  &&&           0.0243 p  &&&                  0.874 p\\
0.0762042     &&&        0.0222 p     &&&               0.801 p    \\
0.0768138    &&&         0.0200 p &&&                 0.730 p  \\
0.0774283    &&&         0.0177 p     &&&               0.652 p \\
0.0777000      &&&       0.0168 p   &&&                0.613 p\\
\hline
\end{tabular}
\end{table*}

\subsection{Inhomogeneity}
\label{sect:inhomo}

Regardless of the cell shape a measure of the inhomogeneity of the inner crust 
is given by what we have called the inhomogeneity factor 
\beqy\label{14}
\Lambda = \frac{1}{V_\mathrm{cell}}\int \left(\frac{n(\pmb{r})}{\bar{n}}
-1\right)^2 d^3\pmb{r} \quad ,
\eeqy
where $V_\mathrm{cell}$ is the cell volume. Of particular interest from the standpoint
of transport properties is the analogous quantity defined entirely in terms
of the proton distribution,
\beqy\label{15}
\Lambda_p = \frac{1}{V_\mathrm{cell}}\int 
\left(\frac{n_p(\pmb{r})}{\bar{n}_p} -1\right)^2 d^3\pmb{r} \quad ,
\eeqy
where $\bar{n}_p = Y_p\bar{n}$. 

Columns 2 and 3 of Table~\ref{tab4} show respectively the values of $\Lambda$ 
for the optimal ETF and ETFSI+BCS spherical configurations, while column 4 
shows the values of $\Lambda$ for the actual equilibrium shape, spherical (s), 
cylinder (c) or plate (p), as the case may be. In columns 5, 6 and 7 
we display the comparable quantities for $\Lambda_p$. The difference in the
value of $\Lambda$ between spherical and pasta shapes never exceeds 6\%, but
for $\Lambda_p$ the difference can amount to 15\%. Thus in this
respect imposing the constraint of a spherical cell shape can do no more than
provide a qualitative guide as to what happens when pasta shapes are allowed.
Nevertheless, it is perhaps remarkable that even this level of similarity
exists, given the quite different cell shapes that are being compared.

Inhomogeneities can be alternatively characterized in terms of the volume 
fraction occupied by clusters. This quantity is of particular interest since,  
in the liquid-drop picture, spherical clusters are predicted to become
unstable against quadrupole deformations when their filling fraction
exceeds $1/8=12.5\%$~\cite{pet95}. Then spaghetti configurations become stable 
in place of spherical ones. At a filling 
fraction of $1/2$ clusters are predicted to ``turn inside out''~\cite{bbp71}. 
On the other hand, quantum-molecular dynamics simulations~\cite{wat09} 
indicated that clusters remain quasi-spherical until they touch similarly to
percolating networks, as speculated earlier by Ogasawara and
Sato~\cite{oga82}. According to these simulations, the onset of pasta
formation is essentially determined by the maximum packing fraction of 
spherical clusters, which is given by $\sqrt{3}\pi/8\simeq 68.0\%$ for 
a body-centered cubic lattice. To compare with these predictions, we have 
estimated the filling fraction of spherical clusters. A natural definition is 
to take $(C_q/R)^3$, depending on whether clusters are characterized by the 
proton or neutron distributions (the definition can be easily extended to 
cylinders and plates, the exponent $3$ being replaced by $2$ and $1$, 
respectively). However, this definition is insensitive to the diffusivity 
coefficients $a_q$ of the nucleon distributions.
As an alternative definition we therefore assume that the nucleons 
characterizing the clusters are uniformly distributed, as in the liquid-drop 
picture, and define the filling fraction in terms of the ratio
\beqy
\frac{V_{\mathrm{cl}}}{V}=\frac{A_{\mathrm{cl}}}{A} \frac{\bar n}{n_{\Lambda,p}+n_{\Lambda,n}}\quad ,
\eeqy
with 
\beqy
A_\mathrm{cl} = 4\pi\,\int_0^R r^2\left[f_n(r)n_{\Lambda n} +f_{p}(r)n_{\Lambda p}\right] \,\mathrm{d}r 
\eeqy
determining the number of nucleons contained in a single cluster. 
These different definitions of the filling fraction lead to remarkably similar
numerical values. More importantly, the threshold value of the filling fraction
above which spherical cells become unstable against pasta formation turns out 
to be in close agreement with the liquid-drop criterion, as can be seen in 
Table~\ref{tab5}. Likewise, inspection of Table~\ref{tab6} shows that the 
transition from cylinders to plates occurs for a volume fraction $\sim 31\%$, 
comparable to that obtained by Hashimoto et al.~\cite{hash84}. Moreover, the 
absence of inverted configurations from our solutions could have been 
anticipated from the fact that the filling fraction never exceeds $55\%$ (see 
Table~\ref{tab7}); cylindrical tubes or spherical bubbles are only predicted at
filling fractions above $\sim 65\%$. However, the back transition cylinder
$\rightarrow$ sphere is not found in liquid-drop models. 

The disagreement of the quantum-molecular dynamics simulations with our
calculations may stem from the fact that the former were carried out at finite 
temperatures and for a fixed proton fraction $Y_p\sim 0.39$, which is much 
higher than expected in the mantle of neutron stars. 

\begin{table*}
\centering
\caption {Filling fractions for spherical clusters from ETFSI+BCS (ETF) calculations.}
\label{tab5}
\begin{tabular}{|cccc|}
\hline 
$\bar{n}$ & $(C_n/R)^3$ & $(C_p/R)^3$ & $V_{\mathrm{cl}}/V$ \\
\hline
0.049000 & 0.121339 (0.123286)& 0.098333 (0.100585)& 0.124656 (0.126328)\\
0.050000 & 0.125640 (0.127702)& 0.102046 (0.104407)& 0.128922 (0.130704)\\
0.051000 & 0.130075 (0.132241)& 0.105898 (0.108365)& 0.133306 (0.135192)\\
0.052000 & 0.134665 (0.136946)& 0.109902 (0.112491)& 0.137826 (0.139832)\\
0.054000 & 0.144367 (0.147488)& 0.118443 (0.121963)& 0.147344 (0.150129)\\
\hline
\end{tabular}
\end{table*}

\begin{table}
\centering
\caption {Filling fractions for cylinders from ETF calculations.}
\label{tab6}
\begin{tabular}{|cccc|}
\hline 
$\bar{n}$ & $(C_n/R)^2$ & $(C_p/R)^2$ & $V_{\mathrm{cl}}/V$ \\
\hline
0.049000 & 0.127064 & 0.101491 &  0.128928 \\
0.050000 & 0.131514 & 0.105288 &  0.133315 \\
0.051000 & 0.136202 & 0.109334 &  0.137894 \\
0.052000 & 0.140894 & 0.113343 &  0.142514 \\
0.054000 & 0.151083 & 0.122233 &  0.152432 \\
0.056000 & 0.162127 & 0.131998 &  0.163108 \\
0.058000 & 0.174214 & 0.142831 &  0.174701 \\
0.060000 & 0.187513 & 0.154930 &  0.187341 \\
0.061000 & 0.194750 & 0.161616 &  0.194177 \\
0.062000 & 0.202324 & 0.168647 &  0.201292 \\
0.063000 & 0.210379 & 0.176160 &  0.208812 \\
0.064000 & 0.218882 & 0.184203 &  0.216719 \\
0.065000 & 0.228075 & 0.192952 &  0.225217 \\
0.066000 & 0.239043 & 0.203723 &  0.235369 \\
0.067000 & 0.248148 & 0.212451 &  0.243640 \\
0.068062 & 0.260073 & 0.224208 &  0.254482 \\
0.069144 & 0.272906 & 0.237092 &  0.266115 \\
0.069255 & 0.274510 & 0.238616 &  0.267560 \\
0.069809 & 0.281509 & 0.245700 &  0.273866 \\
0.070368 & 0.289058 & 0.253392 &  0.280659 \\
0.070931 & 0.296728 & 0.261202 &  0.287515 \\
0.071498 & 0.305086 & 0.269872 &  0.295000 \\
0.072070 & 0.313700 & 0.278948 &  0.302700 \\
0.072647 & 0.322679 & 0.288414 &  0.310702 \\
0.073228 & 0.332248 & 0.298555 &  0.319211 \\
\hline
\end{tabular} 
\end{table}

\begin{table}
\centering
\caption {Filling fractions for plates from ETF calculations.}
\label{tab7}
\begin{tabular}{|cccc|}
\hline 
$\bar{n}$ & $C_n/R$ & $C_p/R$  & $V_{\mathrm{cl}}/V$ \\
\hline
0.070931 & 0.383118 & 0.342791 &  0.369547 \\
0.071498 & 0.392846 & 0.352870 &  0.378884 \\
0.072070 & 0.403015 & 0.363550 &  0.388662 \\
0.072647 & 0.413754 & 0.374796 &  0.398942 \\
0.073228 & 0.425104 & 0.386800 &  0.409820 \\
0.073814 & 0.436886 & 0.399368 &  0.421095 \\
0.074404 & 0.449399 & 0.412764 &  0.433053 \\
0.074999 & 0.462759 & 0.427263 &  0.445855 \\
0.075599 & 0.476290 & 0.441886 &  0.458714 \\
0.076204 & 0.491549 & 0.458818 &  0.473379 \\
0.076814 & 0.507352 & 0.476426 &  0.488523 \\
0.077428 & 0.524407 & 0.495647 &  0.504909 \\
0.077700 & 0.535655 & 0.508414 &  0.515874 \\
\hline
\end{tabular}
\end{table}

\section{Concluding remarks}
\label{concl}

Using the nuclear density functional BSk24 we have generalized our neutron-star
calculations of Ref.~\cite{pea18}
to include the possibility of pasta  shapes for the WS cells of the inner
crust, the earlier calculations having  been confined to spherical cells. The
spherical calculations used the ETF  method with shell and pairing corrections
added, but in the  present pasta calculations neither of these corrections was
made. Thus in  comparing our pasta and spherical-cell results we take two
different forms for the latter. Firstly, we compare our pasta results with the
pure ETF  spherical-cell results, i.e., without the corrections, and find  with
increasing density the sequence sphere $\rightarrow$ cylinder  $\rightarrow$
plate of cell shapes before the transition to the homogeneous  core. We do not
find any ``inverted", i.e., bubble-like, configurations,  although we cannot
exclude their existence in a narrow band of densities immediately below the
crust-core transition. The filling fractions associated with the  phase changes
that we find are in remarkable agreement with predictions based  on liquid-drop
considerations, as is the absence of ``inverted'' solutions. 

Given that there are indications that the corrections to the ETF pasta 
calculations might be significantly smaller than for spherical WS cells, it is 
of interest to compare our pasta results with those of the full spherical
calculations, i.e., with the corrections to the latter included. The most 
significant change that emerges on making this comparison is a more complex 
sequence of cell shapes: sphere $\rightarrow$ cylinder $\rightarrow$ sphere 
$\rightarrow$ cylinder $\rightarrow$ plate. This new feature of a ``back
transition" would be completely eliminated only if the neglected 
corrections in our pasta calculations amounted to more than 30 \% of the
correction terms in the spherical-cell calculations. 
However, to be absolutely certain on this point it would be necessary to 
perform full ETFSI+BCS calculations of pasta. 

While there is a certain narrow band of densities in the inner crust 
in which we cannot be sure whether the shape of the WS cell should be 
cylindrical or spherical, perhaps the most significant conclusion to be drawn 
from the present calculations is that in many respects the precise cell shape 
is irrelevant. In particular, we find that to a very good approximation
the EoS and the proton fraction $Y_p$ are the same whether we assume spherical,
cylindrical or plate-like shapes for the WS cells. Thus the global fitting 
formulas developed in Ref.~\cite{pea18} for spherical cells remain valid here 
to the same level of accuracy as before. Here, over the restricted density range
0.05 fm$^{-3}\lesssim\bar{n}<n_\mathrm{cc}$, we have constructed for each of 
the three phase states more accurate fits that well represent the small 
differences between the different phases.

{\bf The very small energy differences that we find between the different cell shapes
are to be contrasted with the much larger differences found in the very recent 
calculations of Schuetrumpf et al~\cite{sch19}. However, their calculations were
performed at fixed values of the proton fraction $Y_p$ much larger than the  
equilibrium values that we have found here.}

Our results are to some extent dependent on the exact form of the 
parametrization (\ref{5}) that we have chosen for the density distributions.
The ideal solution would be to solve the Euler-Lagrange equations, but this is 
computationally impractical in a large-scale investigation of the scope that we 
have undertaken here.

\begin{acknowledgments}
We thank A.K. Dutta for his contributions to early developments of the
code, and are grateful to Noel Martin for private communications. 
The work of J.M.P.{} was supported by NSERC (Canada) and the Fonds de la
Recherche Scientifique (Belgium) under grant No. CDR-J.0115.18. J.M.P. also
thanks the Universit\'e Libre de Bruxelles for the award of a ``Chaire 
internationale" during the tenure of which the present work was completed. 
The work of N.C.{} was supported by Fonds de la Recherche Scientifique
(Belgium) under grant Nos. CDR-J.0115.18 and IISN 4.4502.19. 
The work of A.Y.P.{} was supported by the Russian Science Foundation
grant 19-12-00133.
\end{acknowledgments}

\appendix

\section{Calculation of Coulomb energy}
\label{appA}

To derive Eqs.~(\ref{7}), (\ref{9}) and (\ref{11}) we begin with the general
expression for the total Coulomb energy of a charge distribution 
$n_\mathrm{ch}(\bf{r})$ 
\beqy\label{A1}
E_\mathrm{C} = \frac{1}{2}e^2\int\,n_\mathrm{ch}(\pmb{r})V_\mathrm{C}(\pmb{r})d^3\pmb{r}   \quad  ,
\eeqy
where the Coulomb field $V_c(\pmb{r})$ satisfies Poisson's equation,
\beqy\label{A2}
\nabla^2V_\mathrm{C}(\pmb{r}) = -4\pi\,n_\mathrm{ch}(\pmb{r}) \quad .
\eeqy
In our case the charge distribution satisfies the neutrality condition
\beqy\label{A3}
\int\,n_\mathrm{ch}(\pmb{r})d^3\pmb{r} = 0   \quad .
\eeqy

For spherical cells Eq.~(\ref{A1}) becomes
\beqy\label{A4}
E_\mathrm{C} = 2\pi\,e^2\int_0^R\,n_\mathrm{ch}(r)V_\mathrm{C}(r)r^2dr \quad
\eeqy
Integrating this by parts we have
\beqy\label{A5}
E_\mathrm{C} = -2\pi\,e^2\int_0^R\,u(r)\frac{dV_\mathrm{C}(r)}{dr}dr  \quad ,
\eeqy
where $u(r)$ is given by Eq.~(\ref{8}) and we have made use of the relations
\beqy\label{A6}
u(r=0) = u(r=R) = 0    \quad ,
\eeqy
the latter following from neutrality. But from Eq.~(\ref{A2}) 
\beqy\label{A7}
4\pi\,n_\mathrm{ch}(r) = -\frac{1}{r^2}\frac{d}{dr}\left(r^2\frac{dV_\mathrm{C}(r)}{dr}\right)
\quad ,
\eeqy
whence
\beqy\label{A8}
\frac{dV_\mathrm{C}(r)}{dr} = -\frac{4\pi}{r^2}u(r) \quad .
\eeqy
Eq.~(\ref{7}) follows at once.

For cylindrical cells Eq.~(\ref{A1}) becomes
\beqy\label{A9}
E_\mathrm{C} = \pi\,e^2\int_0^R\,n_\mathrm{ch}(\eta)V_\mathrm{C}(\eta)\eta\,d\eta \quad
\eeqy
Integrating this by parts we have
\beqy\label{A10}
E_\mathrm{C} = -\pi\,e^2\int_0^R\,u(\eta)\frac{dV_\mathrm{C}(\eta)}{d\eta}d\eta  \quad ,
\eeqy
where $u(\eta)$ is given by Eq.~(\ref{10}) and we have made use of the relations
\beqy\label{A11}
u(\eta=0) = u(\eta=R) = 0    \quad .
\eeqy
But from Eq.~(\ref{A2})
\beqy\label{A12}
4\pi\,n_\mathrm{ch}(\eta) = -\frac{1}{\eta}\frac{d}{d\eta}
\left(\eta\frac{dV_\mathrm{C}(\eta)}{d\eta}\right)
\quad ,
\eeqy
whence
\beqy\label{A13}
\frac{dV_\mathrm{C}(\eta)}{d\eta} = -\frac{4\pi}{\eta}u(\eta) \quad .
\eeqy
Eq.~(\ref{9}) follows at once.

For plate-like cells Eq.~(\ref{A1}) becomes
\beqy\label{A14}
E_\mathrm{C} = e^2\int_0^R\,n_\mathrm{ch}(z)V_\mathrm{C}(z)dz \quad
\eeqy
Integrating this by parts we have
\beqy\label{A15}
E_\mathrm{C} = -e^2\int_0^R\,u(z)\frac{dV_\mathrm{C}(z)}{dz}dz  \quad ,
\eeqy
where $u(z)$ is given by Eq.~(\ref{12}) and we have made use of the relations
\beqy\label{A16}
u(z=0) = u(z=R) = 0    \quad .
\eeqy
But from Eq.~(\ref{A2})
\beqy\label{A17}
4\pi\,n_\mathrm{ch}(z) = -\frac{d}{dz} \left(\frac{dV_\mathrm{C}(z)}{dz}\right)
\quad ,
\eeqy
whence
\beqy\label{A18}
\frac{dV_\mathrm{C}(z)}{dz} = -4\pi\,u(z) \quad .
\eeqy
Eq.~(\ref{11}) follows at once.

\end{document}